\documentclass[usenatbib,a4paper,11pt]{article}
\pdfoutput=1 
\usepackage{jcappub}
\makeatletter
\gdef\@fpheader{}
\makeatother
\usepackage{soul}
\usepackage[T1]{fontenc}
\usepackage{subcaption}
\usepackage[dvipsnames]{xcolor}
\usepackage{soul}
\usepackage{hyperref}
\usepackage{CJK}
\usepackage{multirow}
\usepackage{booktabs}

\usepackage{orcidlink}

\title{\boldmath 
External Attention Transformer:
A Robust AI Model for Identifying Initial Eccentricity Signatures 
in Binary Black Hole Events in Simulated Advanced LIGO Data 
}

\author[a*]{Elahe Khalouei\orcidlink{0000-0001-5098-4165},} 
\author[b*]{Cristiano G. Sabiu\orcidlink{0000-0002-5513-5303},}
\author[a]{Hyung Mok Lee\orcidlink{0000-0003-4412-7161},}
\author[c]{A. Gopakumar\orcidlink{0000-0003-4274-4369}}

\affiliation[a]{Astronomy Research Center, Research Institute of Basic Sciences, Seoul National University, 1 Gwanak-ro, Gwanak-gu, Seoul 08826, Korea}
\affiliation[b]{Natural Science Research Institute, University of Seoul, 163 Seoulsiripdae-ro, Dongdaemun-gu, Seoul 02504, Republic of Korea}
\affiliation[c]{
Department of Astronomy and Astrophysics, Tata Institute of Fundamental Research, Mumbai 400005, India}

\emailAdd{e.khalouei1991@gmail.com}
\emailAdd{csabiu@uos.ac.kr}
\emailAdd{hmlee@snu.ac.kr}
\emailAdd{gopu.tifr@gmail.com}

\footnotetext[1]{Corresponding author(s).}

\abstract{
 Initial orbital eccentricities of gravitational wave (GW) events associated with merging 
 binary black holes (BBHs) should provide clues to their formation scenarios, mainly 
  because various BBH formation channels predict 
 distinct eccentricity distributions. However, searching for inspiral GWs from eccentric BBHs is 
 computationally challenging due to sophisticated approaches to model such GW events.
 This ensures that Bayesian parameter estimation methods to characterize such events 
 are computationally daunting. 
 These considerations influenced us to propose a novel approach to identify 
 and characterize eccentric BBH events in the  LIGO-Virgo-KAGRA (LVK) collaboration data sets 
 that leverages external attention transformer models. Employing simulated data that mimic 
 LIGO O4 run, eccentric inspiral events modeled by an effective-one-body numerical-
relativity waveform family, we show the effectiveness of our approach.
By integrating this transformer-based framework with a convolutional neural network (CNN) architecture, we provide efficient way to identify eccentric 
BBH GW events and accurately characterize their source properties.
}

\begin{document}
\maketitle
\flushbottom

\section{Introduction}
\label{sec:intro}

A decade long routine detections of transient gravitational waves
(GWs) from merging stellar mass binary black holes (BBHs) 
by the LIGO-Virgo-KAGRA (LVK) collaboration 
beginning with 
GW150914 have established the era of GW Astronomy \citep{150914}.
This collaboration have so far detected and characterised 
approximately 100 gravitational wave (GW) events during the three concluded observing runs (i.e. O1-O3) \cite{gwtc1,gwtc3-1, gwtc3}.
The ongoing and planned observational runs 
are expected to should  astrophysical evidences for the
dominant formation channel for these BBHs\citep {BBH_fc}.
We note that the LVK consortium is currently in the second half of the fourth observing run, and the detected gravitational wave (GW) candidates are listed on the GraceDB website
\footnote{https://gracedb.ligo.org/superevents/public/O4/}.
The majority of these signals are most likely produced by binary black hole (BBH) mergers, with the black holes inspiraling in quasi-circular orbits.
\\
It turns out that initial orbital eccentricities, component spins, masses, and eccentricities 
of compact binaries provide crucial clues for distinguishing between different formation channels, as they encapsulate information about the formation history of the system \cite{Mandel, Romero-Shaw1}. As the eccentricity of a binary system rapidly decreases due to the emission of gravitational wave radiation \cite{peter}, isolated binary black holes that inspiral in eccentric orbits are expected to have negligible eccentricities by the time they reach the sensitivity range of LVK observations. 
GW emissions from binaries formed through the isolated evolution process are expected to be detected as they inspiral on quasi-circular orbits. In this scenario, two stars evolve around each other with minimal interaction with their environment, eventually forming compact objects and merging within the age of the Universe \cite{Celoria, Mandel,Romero-Shaw1}.
However, in the case of dynamical formation, certain processes can cause the detected eccentricity to be non-negligible.
Dynamical formation occurs in populated environments, such as globular clusters, young star clusters, and galactic nuclei \cite{DiCarlo,Banerjee,Rodriguez2}.
Merger progenitors formed through the dynamical formation channel may be observed with eccentricities 
$e_{10}$ $>$ 0.1 in the LVK sensitive frequency range (above 10 Hz) \cite{Rodriguez, Samsing2, Zevin, Samsing, Mandel,Tagawa}.
\\
Interestingly, 
several detected events, such as GW190521 \cite{gw19}, have been argued to favor non-negligible orbital eccentricities within the LVK sensitive band (above 10 Hz)\cite{gw19-2,gw19-3,gw19-4,Romeroecc2,Romeroecc,Iglesias}. 
In contrast, 
Ref. \citep{Gupte} re-analyzed 57 GW events using from O1 to O3 observing run of LVK collaboration and 
argued that three BBH events should have non-negligible eccentricities at 10Hz.
Interestingly, these authors find 
no  evidence for orbital eccentricity in GW190521.
Further, it was argued recently that a BH-neutron star merger event, namely GW200105, should have 
a residual orbital eccentricity of $\sim 0.15$ around 20 Hz \cite{Morras}.
\\
We note that the LVK collaboration typically employs matched-filtering techniques to 
detect and characterise transient GW events and the underlying template families 
involve compact binaries in quasi-circular orbits \cite{gwtc3}.
Further, parameter estimation pipelines, such as PyCBC \cite{pycbcPE} and Bilby \cite{bilby}, commonly employ Bayesian inference techniques to thoroughly explore the source parameter space 
of coalescing compact binaries consisting of BHs and neutron stars.
However, the inclusion of eccentricity as an additional parameter substantially increases the computational cost.
Ref. \citep{Divyajyoti} shows that using quasi-circular template banks in search pipelines may lead to missing up to 2.2\% of events, those with a log-uniform distribution of eccentricities. Additionally, this approach may introduce biases in the recovered parameters.
As pointed out in Ref. \citep{Favata}, pipelines using quasi-circular GW templates are capable of detecting eccentric compact binaries with an eccentricity $e_{10}<0.1$ at a frequency of 10 Hz.
For systems with larger eccentricities, it is necessary to conduct eccentricity-targeted searches by constructing templates that account for the effects of eccentricity on the waveform 
\cite{Brown,Huerta,Huerta2}.
Rapidly identifying and characterising these eccentric mergers is therefore crucial for population studies and time-critical multimessenger follow-up. In this work we introduce an \emph{External-Attention Transformer} that sidesteps expensive likelihood evaluations and provides real-time estimates of intrinsic BBH parameters such as 
initial eccentricity and source-frame chirp mass directly from  GW spectrogram data.
\par 
The remainder of this paper is structured as follows. Section~\ref{method} details the model architecture and training procedure. Section ~\ref{result} shows the results of our trained network and assesses its performance on realistic data. We conclude in Section ~\ref{sec:conclusions}.

\section{Methods} \label{method}

Machine learning has been applied to GW data analysis. Some models demonstrate that machine learning can significantly reduce computational costs while maintaining sensitivity comparable to traditional GW data analysis methods.
These techniques have been utilized in various tasks, including signal detection \citep{detection1,detection2,detection3}, glitch classification \citep{glitch}, parameter estimation \citep{PE2,PE1,PE3}
, and a range of other applications  (see \citep{other} and references therein).  


\subsection{Neural Network Architecture}
Transformers \citep{transformer1} were originally developed as building blocks for natural language processing (NLP) tasks (e.g., GPT-4 \citep{GPT4}) and subsequently extended to fields such as computer vision \citep{transformer2} and time-series forecasting \citep{transformer3}. Recently, transformer-based models have been utilized in astronomy and cosmology (e.g. \citep{transformer5,transformer4,transformer6}). 
Although transformers are not yet as commonly used in these fields as other deep learning architectures like convolutional neural networks (CNNs) (e.g. in GW field \citep{other} and references therein), they offer distinct advantages over recurrent neural networks (RNNs) such as long short-term memory (LSTM) \citep{LSTM} networks. Notably, transformers excel at capturing long-range dependencies and enable efficient parallel computation, whereas LSTMs inherently rely on sequential data processing.
\\
The attention mechanism  \citep{transformer1} is a fundamental component of transformer models, designed to capture relationships among different data segments or patches. This mechanism consists of queries, keys, and values. Queries act as prompts seeking specific information, keys function as reference indicators to pinpoint relevant responses, and values hold the underlying data to be accessed and processed. During training, the model learns to compare queries against keys to assess the relevance of each value, thereby enabling effective extraction and representation of meaningful relationships within the data. The self-attention matrix is computed using the softmax function of the scaled dot-product between query and key matrices, subsequently multiplied by the value matrix \citep{transformer1}:
\begin{equation}
\mathrm{Attention_{self}} = \mathrm{softmax}\Big(QK^{T} \Big)V,
\end{equation}
\\
Multi-head attention enhances the capability of single-head attention by projecting queries, keys, and values linearly through different learned projection weights, allowing multiple representations to be captured simultaneously during training. Self-attention mechanisms capture internal relationships within each sample individually, they inherently disregard relationships between different samples. 
To address this limitation, the external attention mechanism has been introduced by EANet \citep{external}, defined as \citep{formula}:
\begin{equation}
\mathrm{Attention_{external}} =  \mathrm{Norm}\Big(Q M_{k}^{T}\Big)M_{v},
\end{equation}
where $M_{k}$ and $M_{v}$ are memory units representing key and value matrices, respectively. These parameters are learned independently from samples, enabling external attention to capture relationships across multiple samples and thereby enhancing generalization.

In this work, we employ an external transformer architecture adapted from the AENet classification model \footnote{\url{https://keras.io/examples/vision/eanet/}}, modifying it specifically for our regression task. 
Additionally, we use two convolutional layers with 64 and 128 filters, respectively, each with a kernel size of $3 \times 3$. Each convolutional layer is subsequently followed by batch normalization and max pooling. After these layers, we apply global average pooling and a dense output layer for the regression task.
We also checked fully connected (dense) neural network layers as an alternative to the CNN, but found that the CNN architecture yielded better performance for our task.
\\
Our goal is to evaluate whether external transformers can effectively identify characteristic patterns associated with orbital eccentricity in BBH  GW events.  
It may be noted that temporally evolving GW polarization states associated with 
 non-spinning BBHs merging along eccentric orbits show features that arise 
from orbital, periastron advance, and GW emission effects as explained in
\cite{DGI}.
We 
generate GW spectrogram samples to create a dataset for training the model and assessing its performance. We then evaluate our model using new unseen mock GW data.

\subsection{Preprocessing: creating the Synthetic Data}\label{sec:data_cosmo}
We employ the SEOBNRv5EHM waveform model \citep{Gamboa,EOB5} from the effective-one-body 
numerical-relativity (EOBNR) waveform model family \citep{EOB1} using the 
$\texttt{pyseobnr}$ package  \citep{pys} and $\texttt{LALSuite}$ \citep{lal1,lal2} to generate simulated eccentric gravitational wave (GW) waveforms.
This model represents an order of improvement over the SEOBNRv4EHM model described in Ref. (\citep{EOB4}, \citep{EOB4-2}) and includes higher-order multipoles: $(\ell,|m|) = (2,2), (2,1), (3,2),
(3,3), (4,3), (4,4)$. For this work, we focus on the dominant $(\ell,|m|) = (2,2)$ mode in our simulations. The chirp mass $M_{c}$ of each binary is drawn uniformly from the range (15, 60) $M_\odot$.
The mass ratio $q = m_2/m_1$ spans the range 0.25 to 1. 
The initial orbital phase $\zeta$ is sampled uniformly over $(0, 2\pi)$, and the eccentricity at 15 Hz, $e_{15,\mathrm{Hz}}$, is drawn uniformly from $(0.001, 0.35)$. Distances are randomly sampled within the range 100 to 1500 Mpc. The sky position (right ascension and declination) and orientation (inclination and polarization angles) are each uniformly sampled over the sky and the polarization sphere, respectively. For simplicity, we neglect the spins of the binary black hole components.
Waveforms are generated at a sample rate of 4096 Hz, with a frequency resolution of 1/16 Hz. Following the methodology outlined in Ref. \citep{adh}, we generate spectrograms of the simulated signals using PyCBC \citep{pycbc}. Each waveform is 30 seconds long, spanning the time window $[-15, 15]$ seconds, and is injected into 40 seconds of Gaussian noise ($[-20, 20]$ seconds) using the O4 sensitivity curve for the LIGO detectors\footnote{\url{https://dcc.ligo.org/ligo-t2000012/public}}. The merger time is set to zero for all synthetic signals.
To reduce computational costs, we consider only the data in the $[-10, 2]$ seconds window for generating spectrograms. We have checked that this time range is sufficient to capture the full signal duration, even for the longest signals in our simulations, which correspond to $M_{c}$=15 $M_\odot$ and q=1.
For each simulated dataset, we calculate the matched-filter signal-to-noise ratio (SNR) and retain only those samples with  LIGO Hanford–LIGO Livingston network SNR greater than 10.
The resulting signals are band-pass filtered in the $[20, 512]$ Hz frequency range and whitened, after which the Q-transform is applied to the whitened signal. The resulting Q-transform spectrograms are resized to $256 \times 256$ pixels and stacked to construct the dataset for subsequent visual processing tasks.
 
To construct our dataset for training and testing the model, we repeat the above simulation procedure to generate 5000 events for each network SNR range: (10–15), (15–20), (20–30), (30–40), (40–50), and (50–60). This results in a total of 30,000 samples, with each sample represented as a spectrogram of shape (256, 256, 2). The three dimensions correspond to frequency, time, and the two LIGO detectors (Hanford and Livingston), respectively.

\subsection{Training}\label{sub2.3}
We divided the simulated GW spectrogram dataset into three subsets: training, testing, and validation. We randomly allocated 80\% of the total sample for training and 20\% for testing. Additionally, 20\% of the training data was set for validation. 
To ensure consistent input distributions, we apply z-score normalization to all data subsets: we first compute the mean and standard deviation on the training set, then subtract the mean and divide by the standard deviation for every sample in the training, validation, and test sets. 
Our model is designed for regression, with a primary focus on eccentricity estimation. However, as shown in Ref. \citep{em,em1}, there exists a degeneracy between chirp mass and eccentricity. To address this, we include both eccentricity and chirp mass as target variables in our model, and both parameters are standardized using the z-score normalization method. During inference, we recover the physical values of the predicted parameters by applying the inverse transformation of the z-score normalization.

\begin{figure*}
\centering
\includegraphics[angle=0,width=0.7\textwidth,clip=]{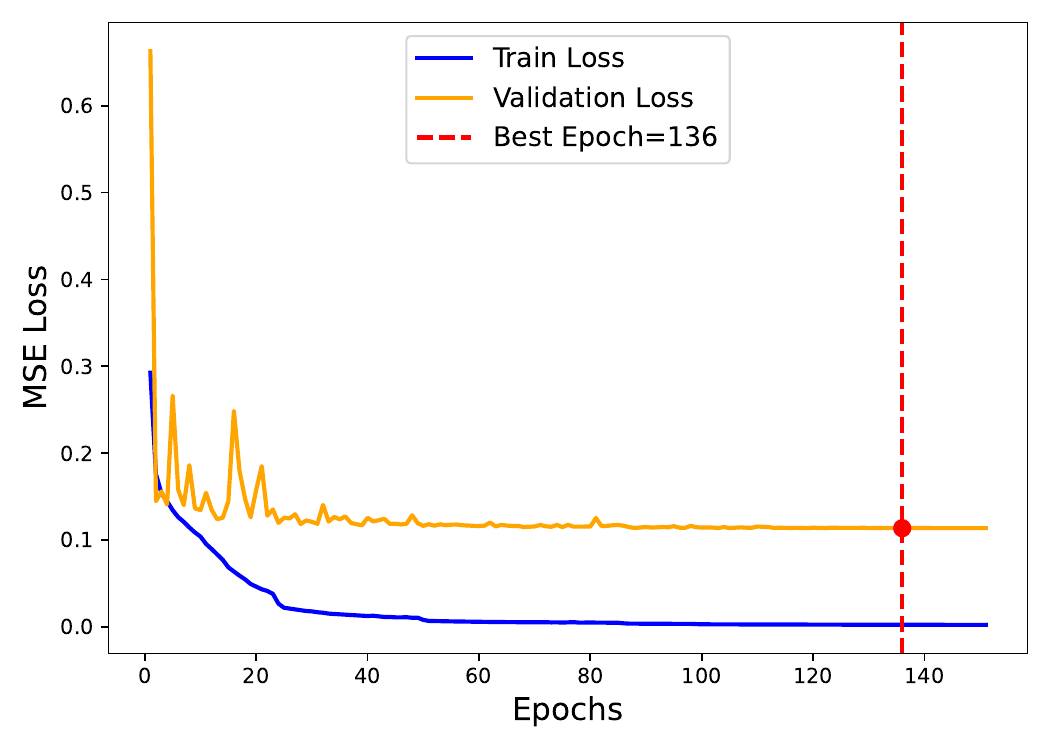}
\caption{ The evolution of MSE loss function during training process. We save and use the model at the epoch with lowest validation loss.
The blue line shows the train loss and the orange line represents the validation loss. The red dashed line marks the epoch where the validation loss reaches its lowest value.
}
\label{loss}
\end{figure*}

\begin{figure*}
\includegraphics[angle=0,width=0.5\textwidth,clip=]{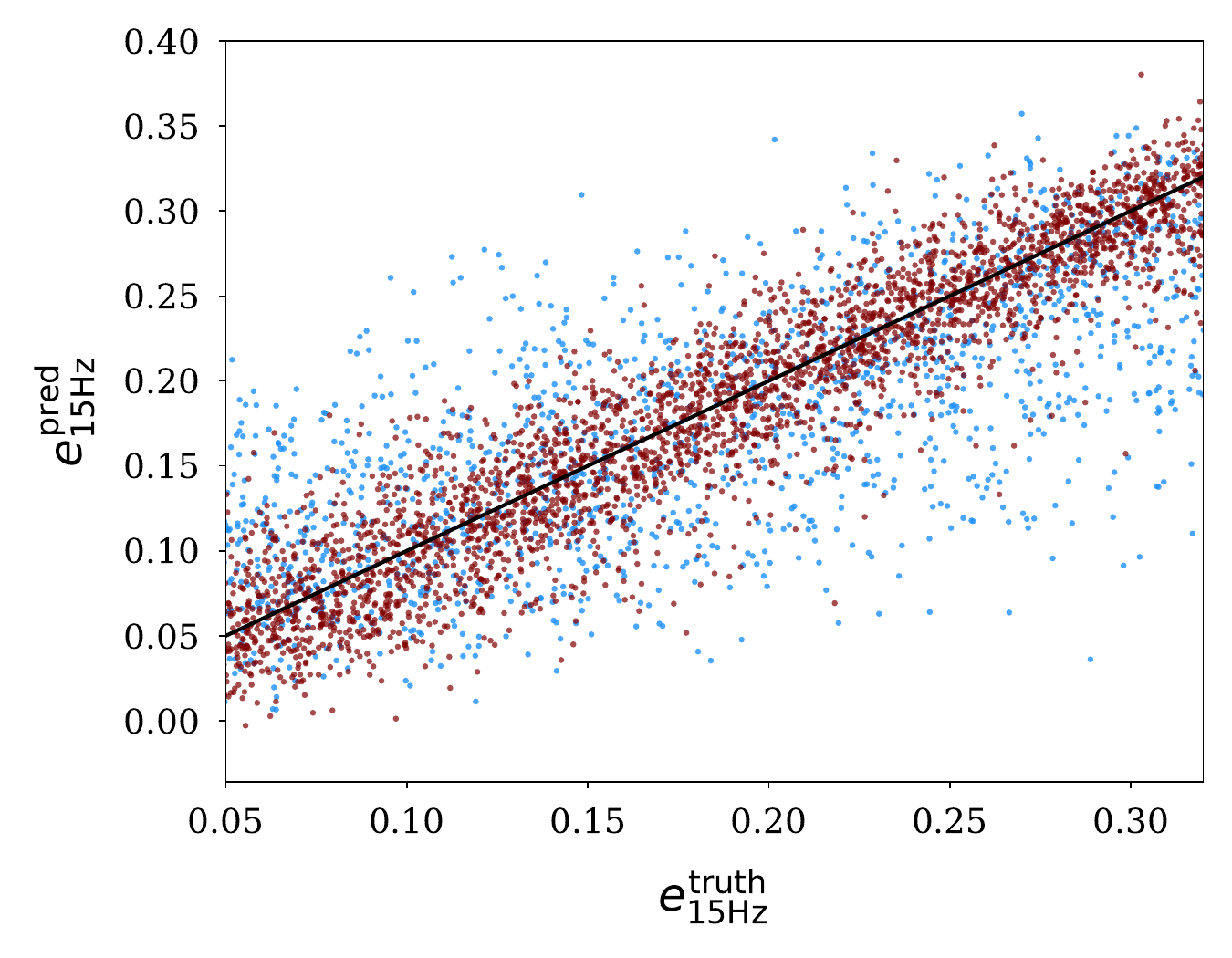}
\includegraphics[angle=0,width=0.5\textwidth,clip=]{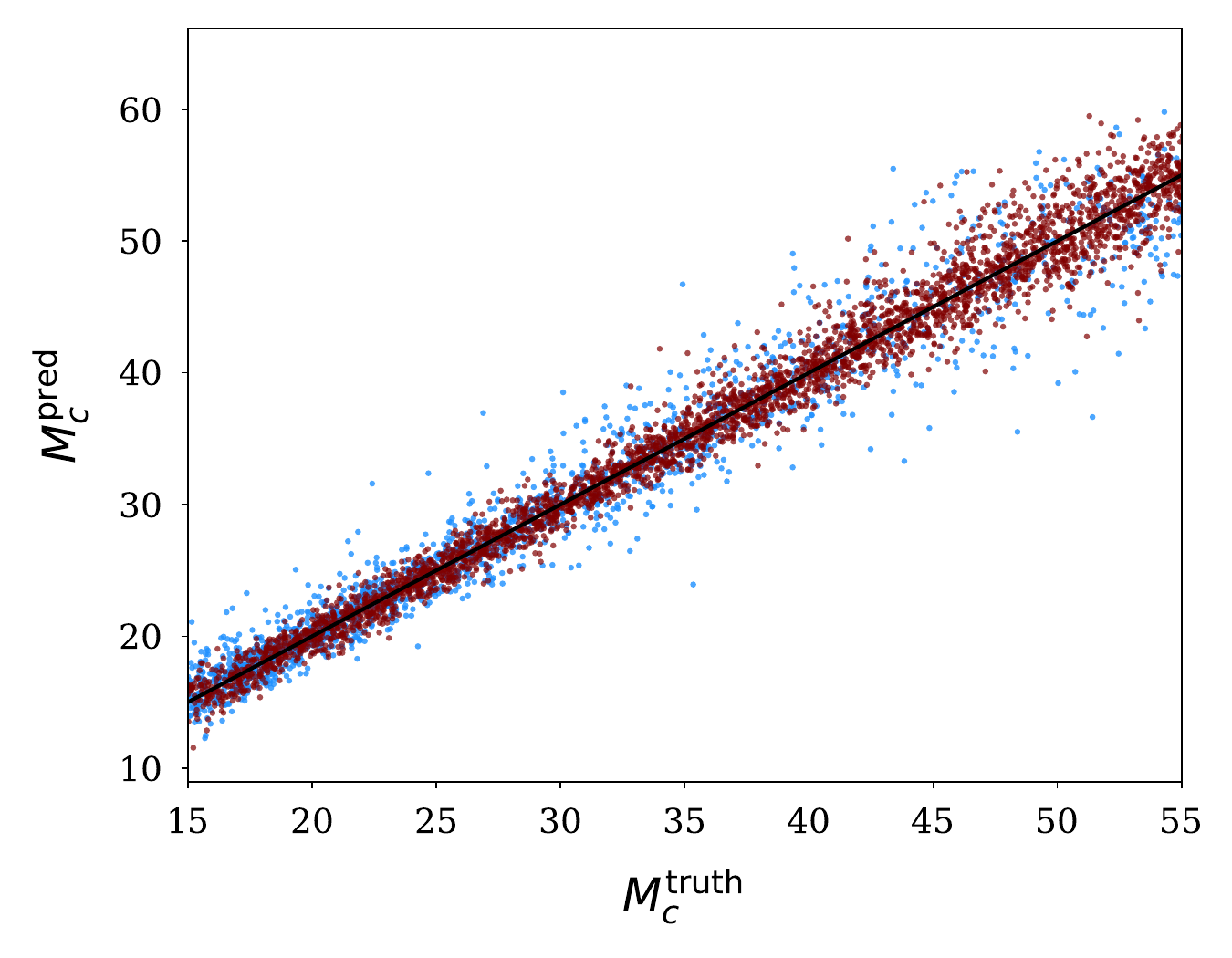}

\caption{A comparison between the input (injected BBH) eccentricity and chirp mass and the corresponding values predicted by the model from the test datasets. Blue points represent  injection samples with 
10 < SNR < 20 and brown points represent samples with 
20 < SNR < 60.}
\label{scat}
\end{figure*}

We fine-tune several hyperparameters of the AENet model 
\footnote{\url{https://keras.io/examples/vision/eanet/}}
to optimize its performance for our specific objectives.
We set the maximum training epochs to 500, implementing early stopping if no improvement in validation loss occurs over 15 consecutive epochs. 
\\
During training, we utilize the Adam optimizer with a learning rate of 0.001, a weight decay of 0.0001, and a batch size of 32. We further configure the training process to reduce the learning rate by a factor of two whenever the validation loss plateaus for 10 epochs. 
Our network divides each spectrogram into smaller patches, each sized $16 \times 16$ pixels. These patches then pass through an embedding layer with a projection dimension of 128, followed by multi-head external attention with eight heads.
\\
The mean squared error (MSE) loss function used for our regression task is defined as:
\begin{equation}
\label{eq1}
LOSS_{(MSE)} = \frac{1}{N_{\rm batch}}\sum_{j=1}^{N_{\rm batch}} \sum_{i=1}^{2} \left( y_{i,j}^{\rm pred} - y_{i,j}^{\rm truth} \right)^2 ~,
\end{equation}
\\
where $y_{1}$ and $y_{2}$ represent the eccentricity and chirp mass, respectively, and $N_{\mathrm{batch}}$ is the number of data in each batch. 
\\
The model training is performed on a single GPU (e.g., NVIDIA GeForce RTX 2070 with 8 GB VRAM) utilizing 80\% of the dataset (approximately 24,000 gravitational wave spectrograms) with an additional 20\% of this subset used for validation. The entire training procedure takes less than one hour.
The model has $1,840,450$ trainable parameters.
\\
Figure \ref{loss} shows the training and validation loss curves as a function of epoch. During training, we monitor the validation loss and select the final model corresponding to the lowest minimum of the validation loss.
Figure \ref{scat} presents scatter plots comparing the model-predicted eccentricity and chirp mass with the true injected values from the test data. 
The Advanced LIGO-Virgo-KAGRA (LVK) detectors are sensitive to orbital eccentricities greater than 0.05 at a frequency of 10 Hz \cite{Lower}. Therefore, we present results only for eccentricities above this threshold.
\\
We perform inference with our trained model for $e_{15,\mathrm{Hz}}$ values in the range [0.05, 0.32] and $M_c$ values in the range [15, 55] $M_\odot$. Although the model is trained on a broader range of these parameters, we restrict the inference range to avoid systematic bias arising from edge effects in the data \citep{mbias}.
In the right panel of Figure \ref{scat}, we observe increased scatter in the recovered $M_c$ values at higher masses. This is expected, as the time-frequency maps become shorter for higher-$M_c$ binary black holes, leading to greater uncertainties in the recovered $M_c$ values with our model.
In the left panel of Figure \ref{scat}, we find that for low/high eccentricity, the model tends to overestimate/underestimate values when the network SNR is low, indicating difficulty in extracting reliable eccentricity information from low-quality data. Therefore, we need to investigate to determine if there are any significant systematic biases in the estimation.
In the next section, we examine the distributions of eccentricity and chirp mass in greater detail. 
 
\section{Implications of Our Approach }
\label{result}
\begin{figure*}[!ht]
    \centering
    \begin{subfigure}[b]{0.45\textwidth}
        \centering
        \includegraphics[width=\textwidth]{ 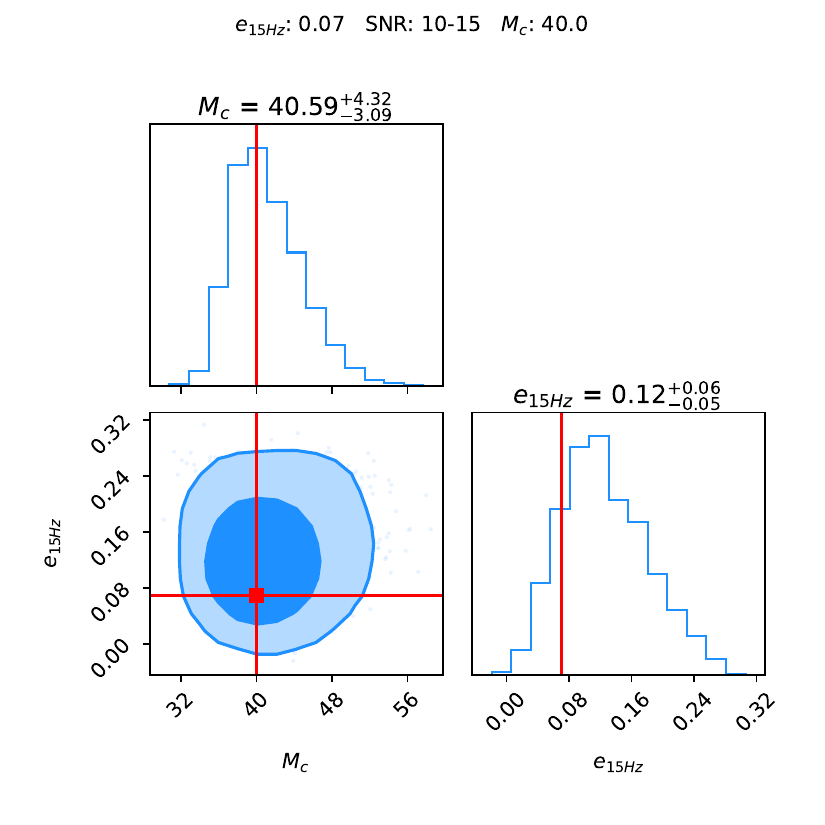}  
    \end{subfigure}
    \hfill
    \begin{subfigure}[b]{0.45\textwidth}
        \centering
        \includegraphics[width=\textwidth]{ 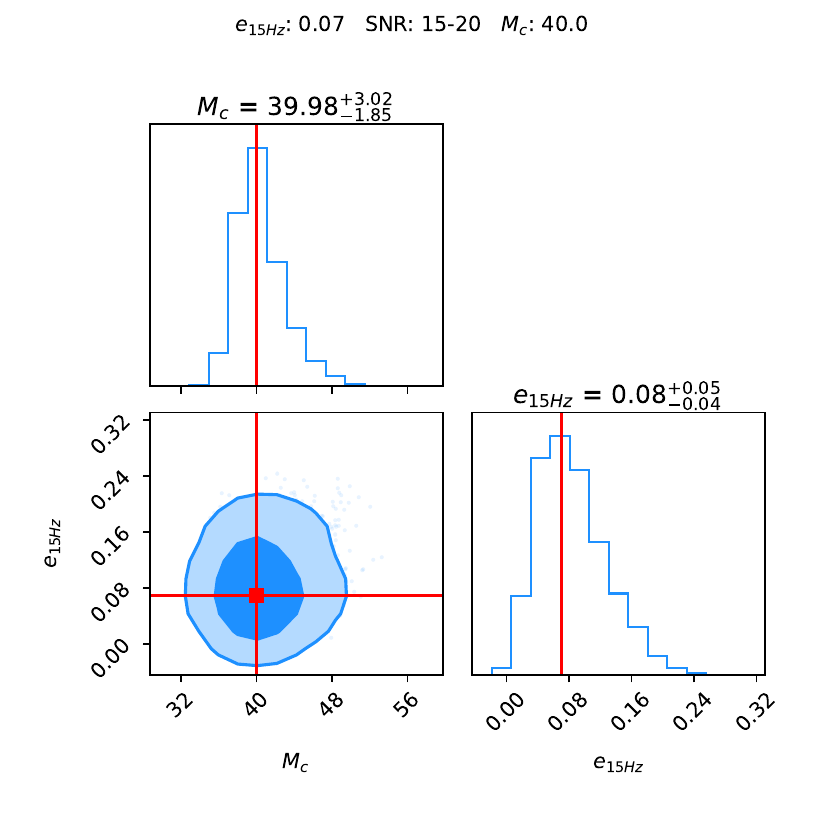}
       
    \end{subfigure}
    \begin{subfigure}[b]{0.45\textwidth}
        \centering
        \includegraphics[width=\textwidth]{ 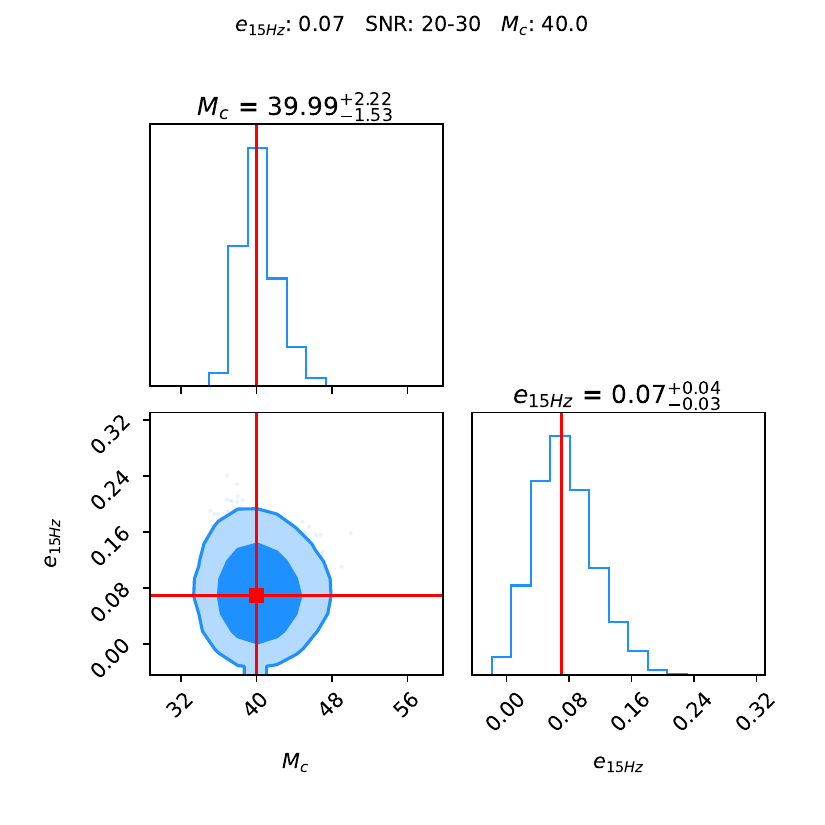}
      
    \end{subfigure}
    \hfill
    \begin{subfigure}[b]{0.45\textwidth}
        \centering
        \includegraphics[width=\textwidth]{ 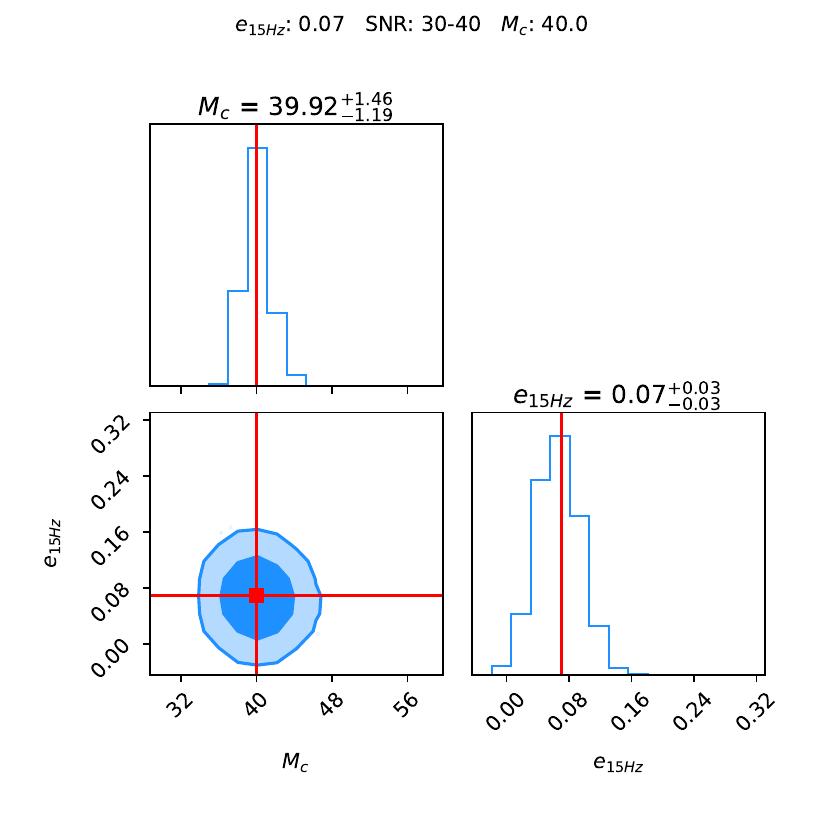}
       
    \end{subfigure}
    \begin{subfigure}[b]{0.45\textwidth}
        \centering
        \includegraphics[width=\textwidth]{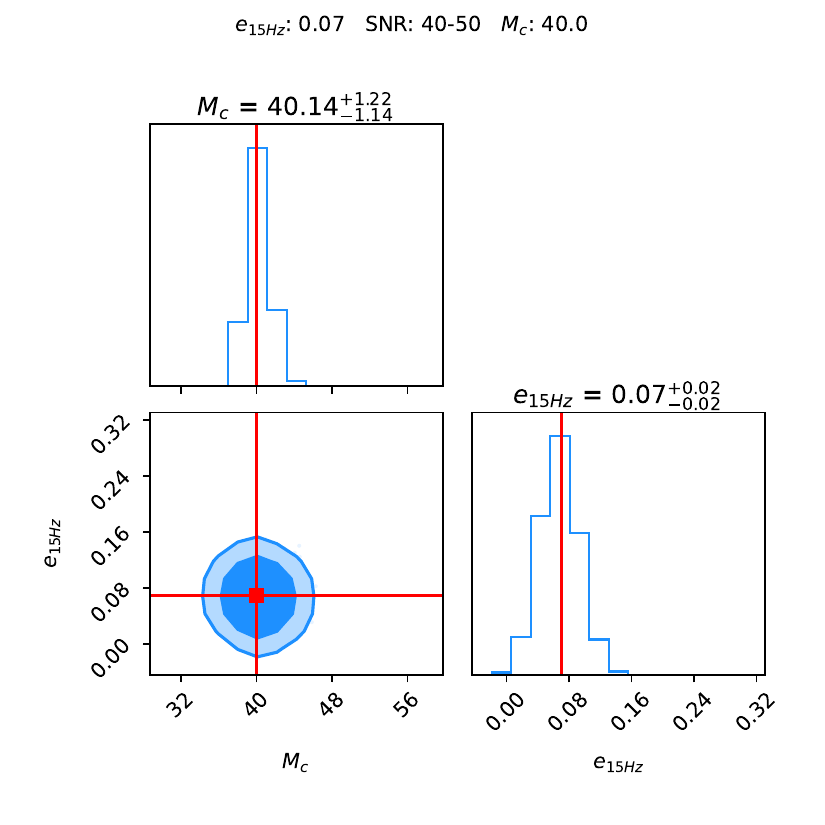}
       
    \end{subfigure}
    \begin{subfigure}[b]{0.45\textwidth}
        \centering
        \includegraphics[width=\textwidth]{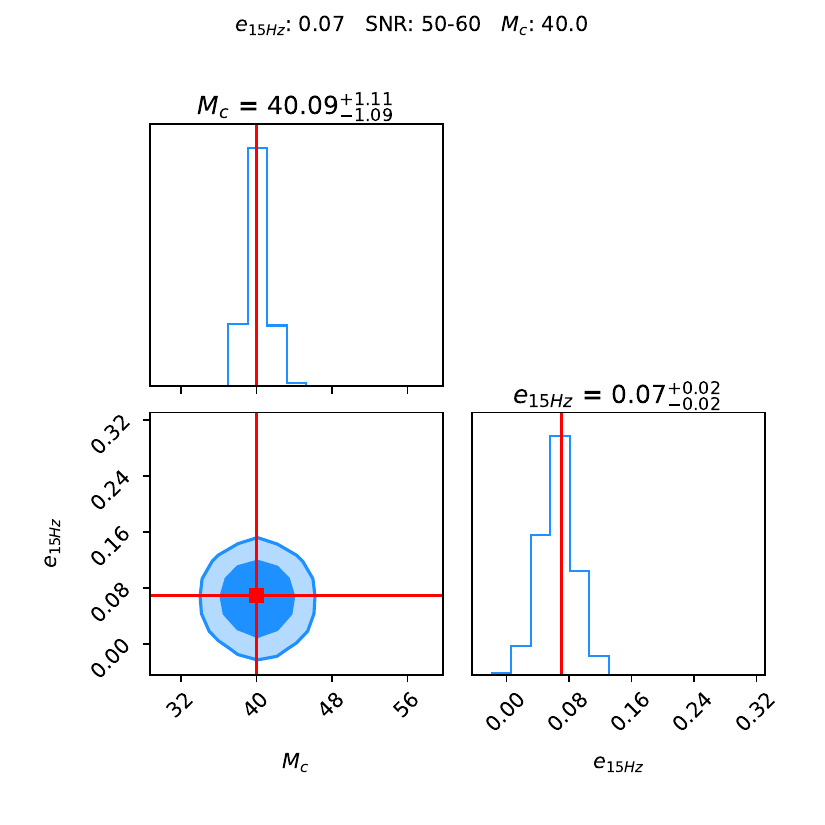}
       
    \end{subfigure}
    \caption{the marginalised 2D contours of eccentricity and chirp mass from prediction of model on several test data sets with fixed eccentricity $e_{15Hz}=0.07$  and chirp mass $M_{c}=40$ $M_\odot$ but different network SNRs. The red lines indicate the injected (true) parameter values. The contours correspond to 1$\sigma$ (inner line) and 2$\sigma$ (outer line) confidence intervals.}
    \label{corner1}
\end{figure*}

\begin{figure*}[!ht]
    \centering
    \begin{subfigure}[b]{0.45\textwidth}
        \centering
        \includegraphics[width=\textwidth]{ 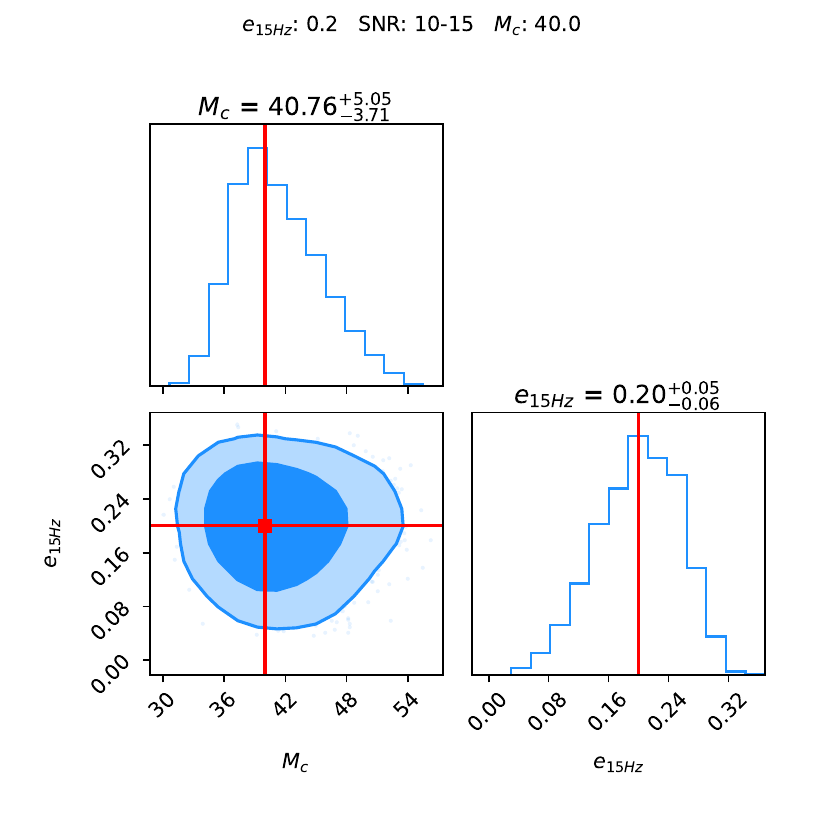}  
    \end{subfigure}
    \hfill
    \begin{subfigure}[b]{0.45\textwidth}
        \centering
        \includegraphics[width=\textwidth]{ 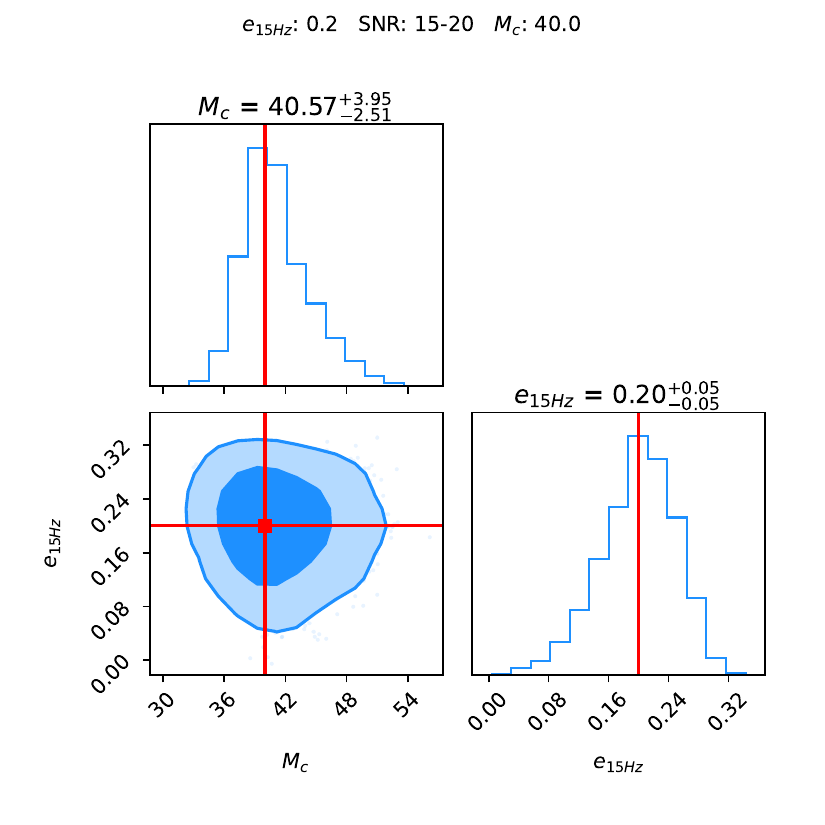}
       
    \end{subfigure}
    \begin{subfigure}[b]{0.45\textwidth}
        \centering
        \includegraphics[width=\textwidth]{ 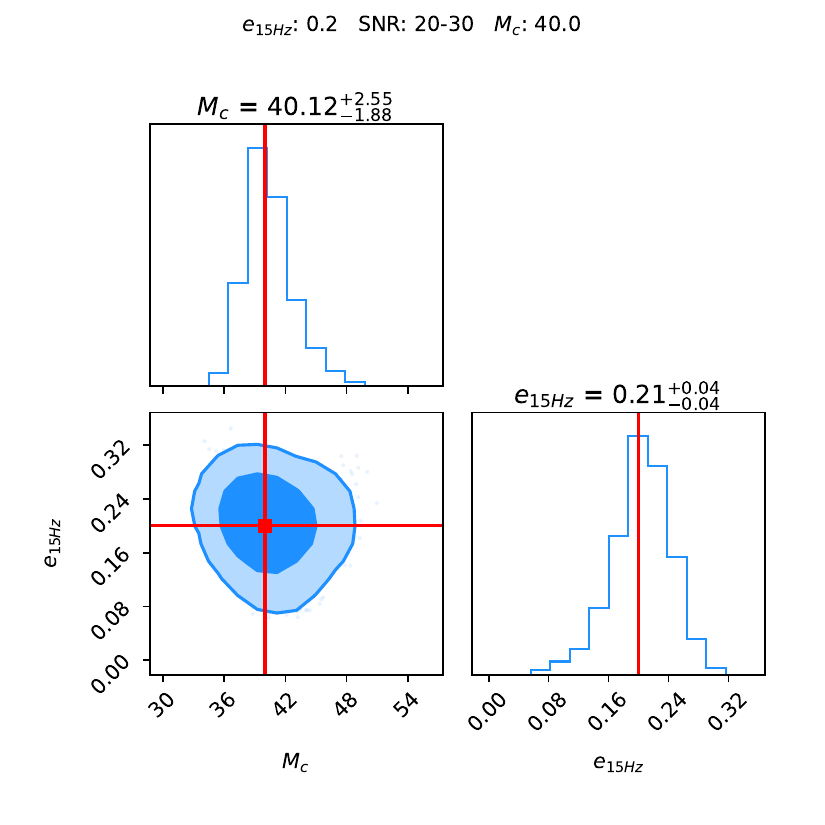}
      
    \end{subfigure}
    \hfill
    \begin{subfigure}[b]{0.45\textwidth}
        \centering
        \includegraphics[width=\textwidth]{ 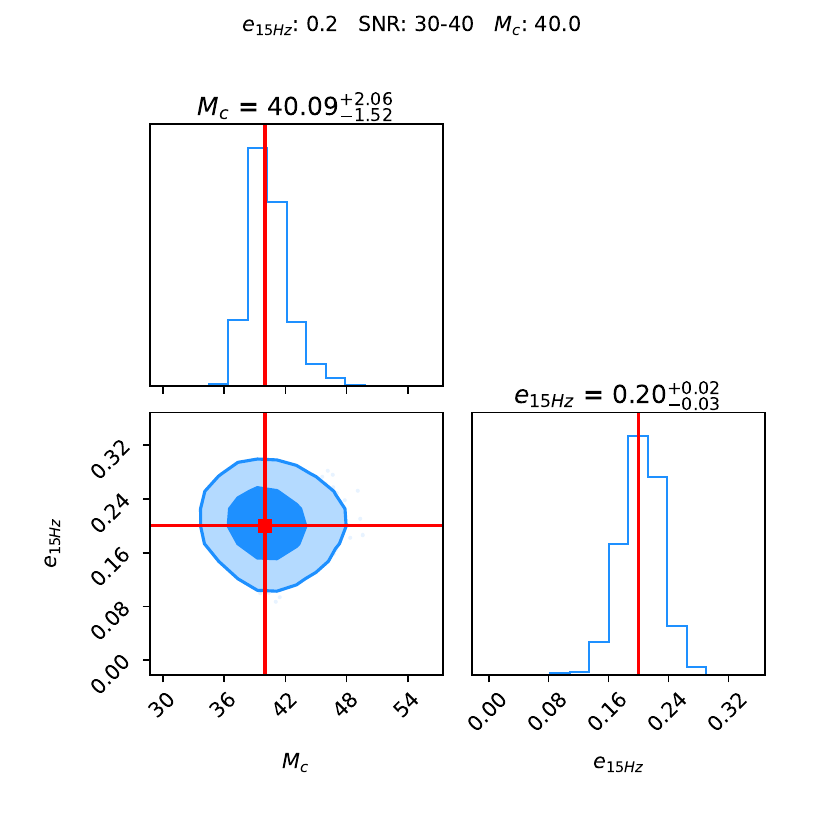}
       
    \end{subfigure}
    \begin{subfigure}[b]{0.45\textwidth}
        \centering
        \includegraphics[width=\textwidth]{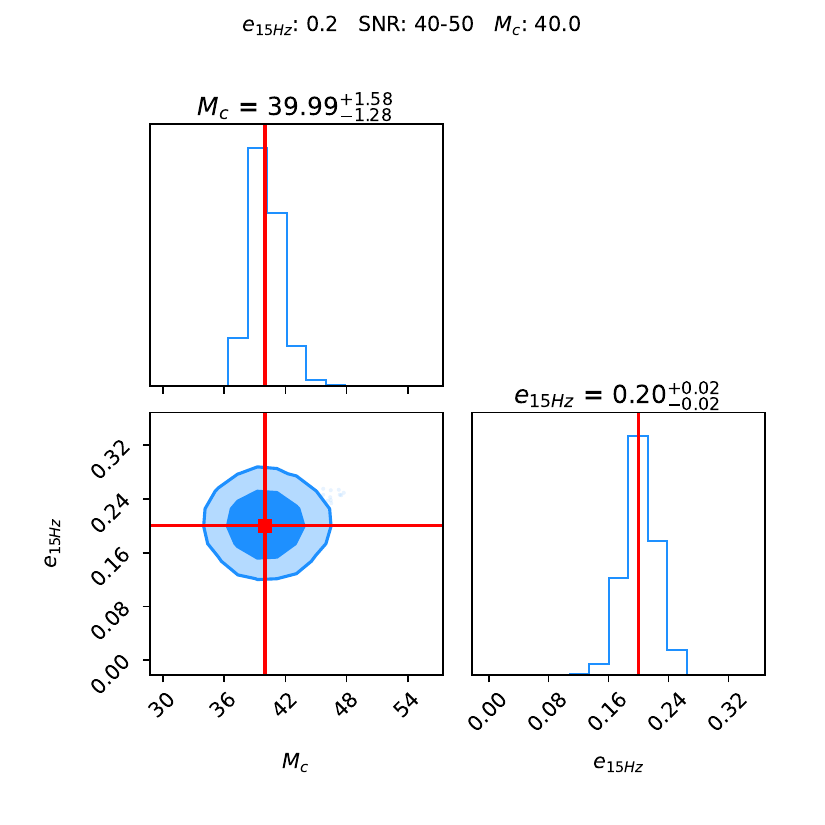}
       
    \end{subfigure}
     \begin{subfigure}[b]{0.45\textwidth}
        \centering
        \includegraphics[width=\textwidth]{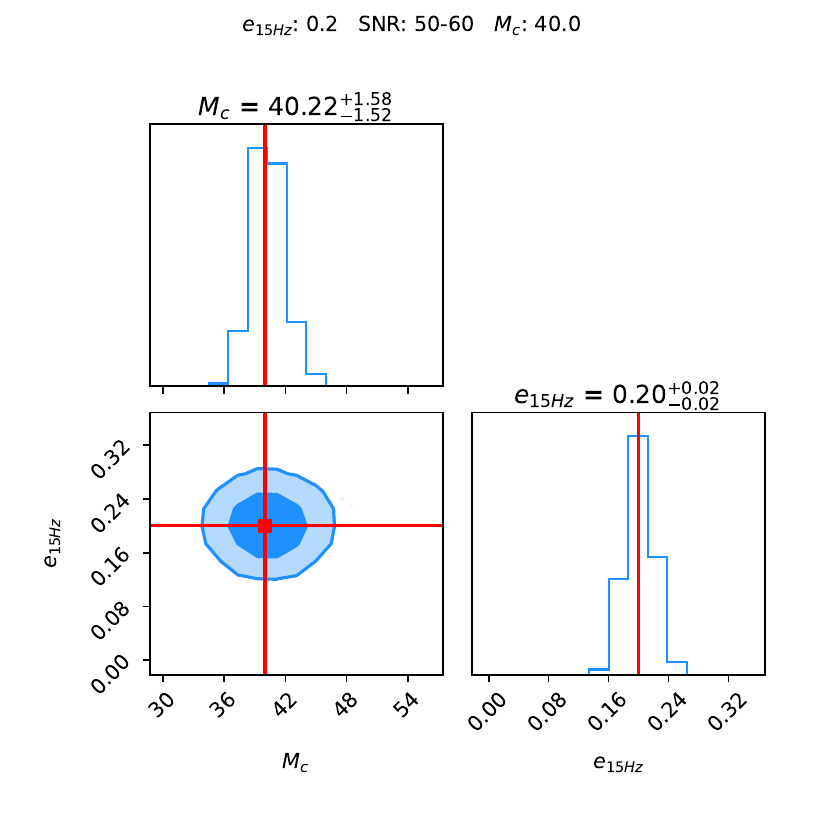}
       
    \end{subfigure}
    \caption{the marginalised 2D contours of eccentricity and chirp mass from prediction of model on several test data sets with fixed eccentricity $e_{15Hz}=0.2$  and chirp mass $M_{c}=40$ $M_\odot$ but different network SNRs. The red lines indicate the injected (true) parameter values. The contours correspond to 1$\sigma$ (inner line) and 2$\sigma$ (outer line) confidence intervals.}
    \label{corner2}
\end{figure*}

\begin{figure*}[!ht]
    \centering
    \begin{subfigure}[b]{0.45\textwidth}
        \centering
        \includegraphics[width=\textwidth]{ 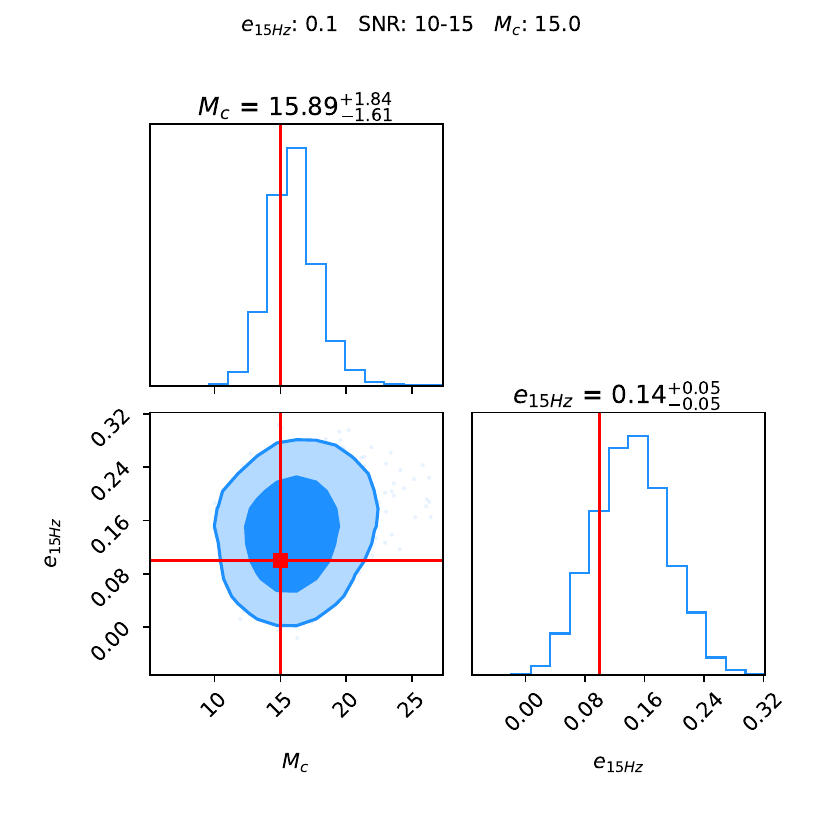}  
    \end{subfigure}
    \hfill
    \begin{subfigure}[b]{0.45\textwidth}
        \centering
        \includegraphics[width=\textwidth]{ 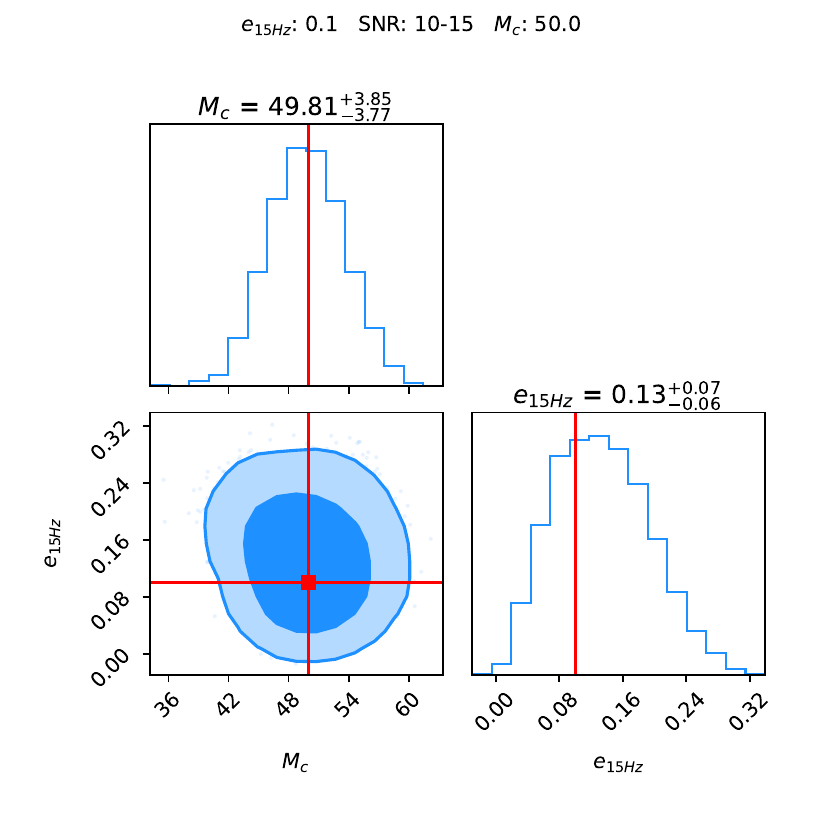}
       
    \end{subfigure}
    \caption{the marginalised 2D contours of eccentricity and chirp mass from prediction of model on several test data sets with fixed eccentricity $e_{15Hz}=[0.10]$, chirp mass $M_{c}=$[15,50] $M_\odot$, and network SNRs=(10-15). The red lines indicate the injected (true) parameter values. The contours correspond to 1$\sigma$ (inner line) and 2$\sigma$ (outer line) confidence intervals.}
    \label{corner5}
\end{figure*}

\begin{figure*}[!ht]
    \centering
    \begin{subfigure}[b]{0.45\textwidth}
        \centering
        \includegraphics[width=\textwidth]{ 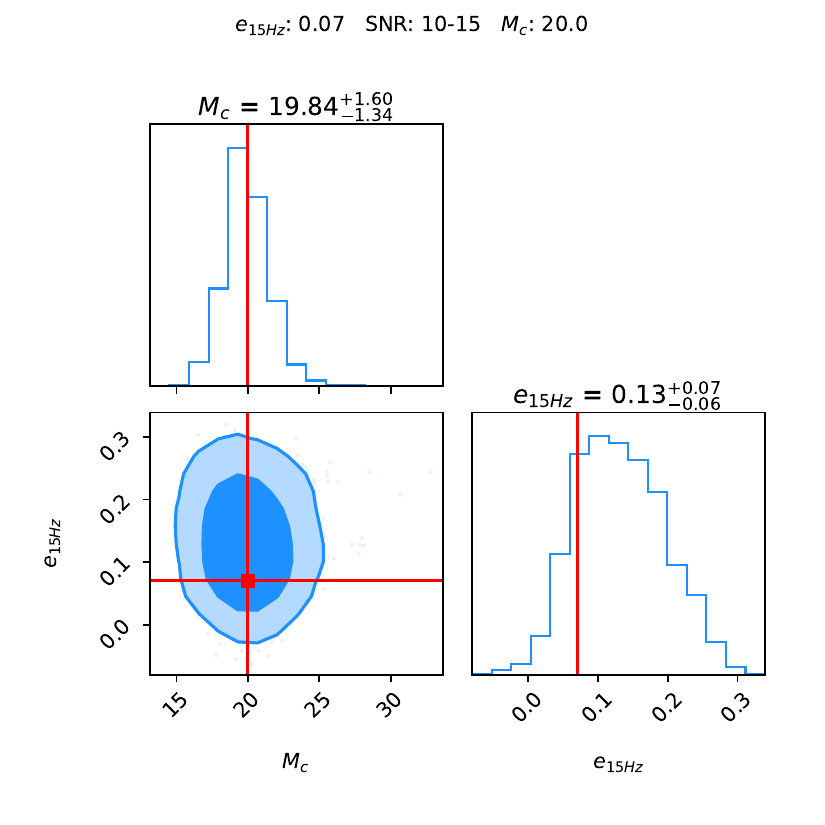}  
    \end{subfigure}
    \hfill
    \begin{subfigure}[b]{0.45\textwidth}
        \centering
        \includegraphics[width=\textwidth]{ 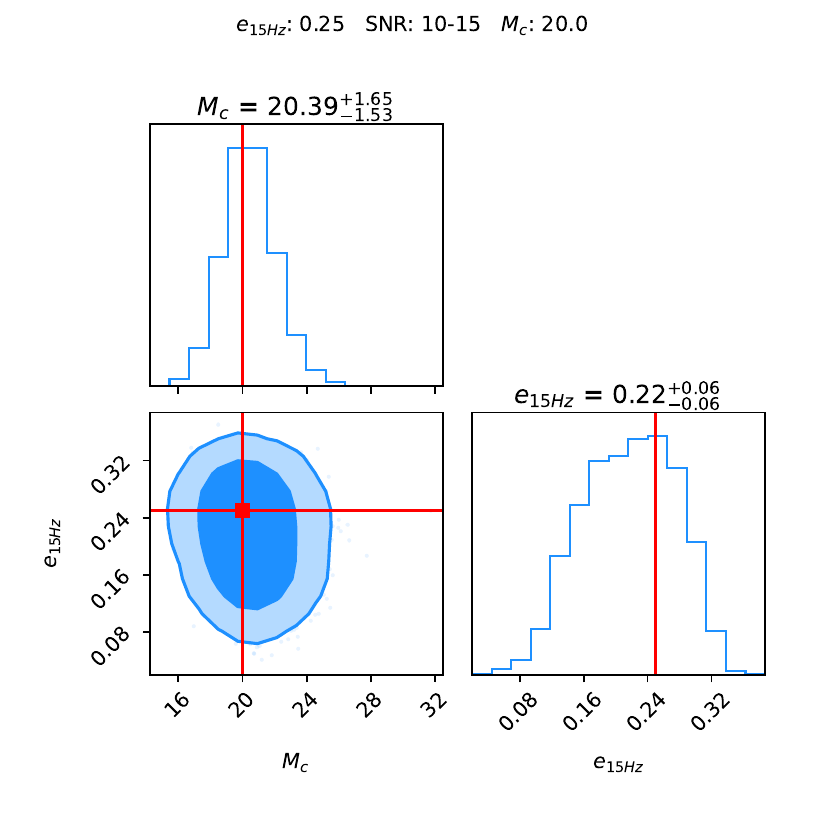}
       
    \end{subfigure}
    \caption{the marginalised 2D contours of eccentricity and chirp mass from prediction of model on several test data sets with fixed eccentricity $e_{15Hz}=[0.07,0.25]$, chirp mass $M_{c}=$[20] $M_\odot$, and network SNRs=(10-15). The red lines indicate the injected (true) parameter values. The contours correspond to 1$\sigma$ (inner line) and 2$\sigma$ (outer line) confidence intervals.}
    \label{corner6}
\end{figure*}

To evaluate the performance of our model, we generate several independent test datasets, each with fixed values of chirp mass and eccentricity, across different network SNR ranges: (10–15), (15–20), (20–30), (30–40), (40–50), and (50–60). For each network SNR range, we generate 300 events. On average, the model required on the order of 100 ms to process a single event, as determined from 100 repeated inferences.
We trained our model independently 10 times, and for each run, we saved the best-performing model based on the minimum validation loss, following the procedure described in subsection \ref{sub2.3}.
Figure \ref{corner1} shows the recovered distributions of $e_{15,\mathrm{Hz}}$ = 0.07 and $M_{c}$ = 40 $M_\odot$ for 300 events in each network SNR bin. In each subpanel, the parameters are recovered within 1 $\sigma$, and the size of the contours decreases as the network SNR increases, indicating improved precision.
Figure \ref{corner2} presents a similar analysis for a higher eccentricity value of 0.2.
\\
In Figure \ref{corner5}, we present the recovered distributions of parameters for low and high chirp masses ($M_c = 15$ and $50 M_\odot$), with eccentricity fixed at $e_{15,\mathrm{Hz}} = 0.10$ and network SNR in the range $10 < \mathrm{SNR} < 15$,
to further investigate the increased scatter observed at higher $M_c$ values in Figure \ref{scat}. We find that the recovered values remain within the 1$\sigma$ region, indicating that the model does not exhibit significant systematic bias in this regime.
\\
Similarly, in Figure \ref{corner6}, we show the recovered parameters for low and high eccentricities ($e_{15,\mathrm{Hz}} = 0.07$ and $0.25$), with the chirp mass fixed at $M_c = 20 M_\odot$ and network SNR in the range $10 < \mathrm{SNR} < 15$. The results demonstrate that the recovered $e_{15,\mathrm{Hz}}$ values are also within the 1$\sigma$ region, suggesting robust performance of the model even at low SNR for both low and high eccentricities.\\
In the corner plots, the 68\% (1$\sigma$) and 95\% (2$\sigma$) contours are empirical. We fix the injected $e_{15,\mathrm{Hz}}$ and $\mathcal{M}_c$ and allow other parameters and the noise realization to vary across events. We then plot the smallest regions enclosing 68\% and 95\% of predictions while noting that these are not formal predictive uncertainty intervals. They summarize dispersion from noise and variation in non-target parameters (e.g., luminosity distance, sky position, inclination) and quantify robustness.
The recovered $e_{15,\mathrm{Hz}}$ and  $\mathcal{M}_c$ values lie within 1$\sigma$ in all SNR bins.  
\\

To better characterize prediction confidence in Figures \ref{corner1}-\ref{corner6}, we evaluate model uncertainty using a deep-ensemble approach. We performed inference using all 10 trained models on several datasets and computed the standard deviation of the predictions for each event. We find that when the SNR is low, the main issue is the strong noise in the data. This noise makes it difficult for the model to identify true patterns, leading to greater variation in its predictions. In other words, most of the uncertainty at low SNR is due to the data, not limitations of the model. As the SNR increases and the signal becomes more prominent relative to the noise, the model's predictions become much more precise and confident. In this high SNR regime, model uncertainty is comparable to the total uncertainty. The ensemble shows no significant improvement over a single model.
\\

\subsection{Quantitative Performance Metrics}
\label{sec:quantitative-metrics}
To further quantify model performance, we report the mean absolute error (MAE), root mean squared error (RMSE), and their associated 95\% confidence intervals across bins of $e_{15,\mathrm{Hz}}$ and $M_{c}$, separately for low and high SNR categories. 
For this evaluation, we generated a new test dataset, with the data binned by $e_{15,\mathrm{Hz}}$, $M_{c}$ , and SNR using the following bin edges:
$e_{15,\mathrm{Hz}}$ bins of (0.05–0.10), (0.10–0.15), (0.15–0.20), (0.20–0.25), and (0.25–0.30); $M_{c}$ bins of (15–25), (25–35), (35–45), and (45–55) $M_{\odot}$; and SNR bins of (10–15), (15–20), (20–30), (30–40), (40–50), and (50–60). 
We used $N = 100$ test samples in each $(e_{15\,\mathrm{Hz}},\, M_c,\, \mathrm{SNR})$ bin combination.
Throughout this analysis, we define low SNR as $10 < SNR < 20$ and high SNR as $20 < SNR < 60$.
The detailed results are summarized in Table~\ref{tab:mae-rmse-ecc} for $e_{15,\mathrm{Hz}}$ bins and Table~\ref{tab:mae-rmse-mc} for $M_{c}$ bins. These quantitative metrics provide a rigorous assessment of the model’s predictive accuracy across the relevant parameter space.  
As expected, the model achieves better performance for lower $M_{c}$ systems, whereas the errors increase for higher $M_{c}$. This trend, discussed in Section \ref{sub2.3}, arises because the time-frequency maps become shorter for higher $M_{c}$, leading to greater uncertainties in the recovered $M_{c}$. For $e_{15,\mathrm{Hz}}$ the model shows consistently reliable performance across the explored range.

\renewcommand{\arraystretch}{1.4}
\begin{table}[ht]
\centering
\caption{MAE and RMSE for $e_{15,\mathrm{Hz}}$ bins with 95\% confidence intervals over ten independent runs.}
\begin{tabular}{c|cc|cc}
\hline
$e_{15,\mathrm{Hz}}$  & MAE$_{\rm SNR_{low}}$ & RMSE$_{\rm SNR_{low}}$ & MAE$_{\rm SNR_{high}}$ & RMSE$_{\rm SNR_{high}}$ \\
\hline
0.05--0.10 & $0.0490^{+0.0027}_{-0.0032}$ & $0.0637^{+0.0033}_{-0.0029}$ & $0.0216^{+0.0009}_{-0.0010}$ & $0.0281^{+0.0015}_{-0.0012}$ \\
0.10--0.15 & $0.0424^{+0.0022}_{-0.0021}$ & $0.0529^{+0.0025}_{-0.0026}$ & $0.0244^{+0.0011}_{-0.0011}$ & $0.0316^{+0.0015}_{-0.0013}$ \\
0.15--0.20 & $0.0422^{+0.0020}_{-0.0020}$ & $0.0519^{+0.0022}_{-0.0024}$ & $0.0247^{+0.0012}_{-0.0011}$ & $0.0323^{+0.0017}_{-0.0016}$ \\
0.20--0.25 & $0.0421^{+0.0023}_{-0.0023}$ & $0.0531^{+0.0026}_{-0.0025}$ & $0.0206^{+0.0009}_{-0.0010}$ & $0.0265^{+0.0013}_{-0.0013}$ \\
0.25--0.30 & $0.0415^{+0.0023}_{-0.0025}$ & $0.0535^{+0.0027}_{-0.0032}$ & $0.0181^{+0.0011}_{-0.0008}$ & $0.0251^{+0.0018}_{-0.0019}$ \\
\hline
\end{tabular}
\label{tab:mae-rmse-ecc}
\end{table}

\begin{table}[ht]
\centering
\caption{MAE and RMSE for $\mathcal{M}_c$  bins with 95\% confidence intervals over ten independent runs. }
\begin{tabular}{c|cc|cc}
\hline
$\mathcal{M}_c$ ($M_{\odot}$) & MAE$_{\rm SNR_{low}}$ & RMSE$_{\rm SNR_{low}}$ & MAE$_{\rm SNR_{high}}$ & RMSE$_{\rm SNR_{high}}$ \\
\hline
15--25   & $0.8002^{+0.0500}_{-0.0460}$ & $1.1008^{+0.1333}_{-0.0995}$ & $0.4813^{+0.0199}_{-0.0184}$ & $0.6084^{+0.0208}_{-0.0210}$ \\
25--35   & $1.3042^{+0.0871}_{-0.0776}$ & $1.8262^{+0.1346}_{-0.1316}$ & $0.6030^{+0.0257}_{-0.0268}$ & $0.8116^{+0.0636}_{-0.0556}$ \\
35--45   & $2.4309^{+0.1292}_{-0.1289}$ & $3.2095^{+0.1688}_{-0.1703}$ & $1.1561^{+0.0607}_{-0.0655}$ & $1.6413^{+0.0921}_{-0.0912}$ \\
45--55   & $2.8756^{+0.1432}_{-0.1326}$ & $3.6514^{+0.1905}_{-0.1673}$ & $1.6351^{+0.0773}_{-0.0662}$ & $2.1609^{+0.1087}_{-0.0938}$ \\
\hline
\end{tabular}
\label{tab:mae-rmse-mc}
\end{table}

\section{Conclusions}
\label{sec:conclusions}
Most LVK pipeline searches currently neglect the eccentricity parameter in gravitational wave (GW) signals due to the significant computational overhead associated with incorporating this parameter into Bayesian inference methods for parameter estimation. Additionally, waveform models accounting for eccentricity are computationally intensive and relatively slow. Consequently, standard analysis pipelines typically assume quasi-circular orbits for parameter estimation. However, neglecting eccentricity in the parameter estimation of genuinely eccentric binary black hole (BBH) events can introduce biases in the recovered parameters \citep{Divyajyoti}.

In this study, we utilize an external transformer-based neural network to effectively identify the eccentricity of BBH systems from GW spectrogram data. Our approach modifies the external transformer classification model, initially introduced by Ref. \citep{external}, converting it into a regression model with the addition of convolutional neural network (CNN) layers. This enhanced model successfully estimates both chirp mass and eccentricity directly from GW spectrograms.
\\
A significant advantage of our method is its rapid and computationally efficient training process, enabled by the external attention mechanism, which employs external memory to capture general patterns effectively across subsamples. 
Specifically, we trained the model on 24,000 gravitational wave spectrograms, corresponding to the training dataset, using a single GPU (NVIDIA GeForce RTX 2070 with 8 GB VRAM). During training, 20\% of the training data was used for validation, and the entire training process was completed in less than one hour.
The average inference time per event, measured over 100 runs, was on the order of 100 ms.
We implemented ensemble learning to quantify model uncertainty. Testing across diverse datasets, as discussed in Section \ref{result}, demonstrates that our model achieves sufficient precision in determining eccentricity.
Our findings show that the ensemble method does not improve performance compared to the single model.
Recognizing the intrinsic degeneracy between eccentricity and chirp mass, our approach simultaneously estimates both parameters to improve robustness and reliability. Furthermore, the model is trained on data simulating the current LVK O4 observational run, making it immediately applicable to ongoing analyses. Due to its efficient training capability, this model can easily be retrained or updated for future LVK observational runs, highlighting its practical utility for rapid and accurate GW parameter estimation.
These considerations imply that our approach should allow us to provide independent constraints on the reported initial eccentricities of a few O3 GW events \citep{gw19,Gupte,Morras}. 
This is currently under investigation. 

\acknowledgments
E.K. and H.M.L. are supported by the National Research Foundation of Korea 2021M3F7A1082056. 
C.G.S. acknowledge support from the Basic Science Research Program (2018R1A6A1A06024977 and RS-2025-00515276, respectively) through Korea's NRF funded by the Ministry of Education.

\bibliographystyle{JHEP}
\bibliography{ref}

\providecommand{\href}[2]{#2}\begingroup\raggedright\begin{thebibliography}{10}

\bibitem{150914}
B.P.~{Abbott}, R.~{Abbott}, T.D.~{Abbott}, M.R.~{Abernathy}, F.~{Acernese}, K.~{Ackley} et~al., \emph{{Observation of Gravitational Waves from a Binary Black Hole Merger}}, \href{https://doi.org/10.1103/PhysRevLett.116.061102}{\emph{\prl} {\bfseries 116} (2016) 061102} [\href{https://arxiv.org/abs/1602.03837}{{\ttfamily 1602.03837}}].

\bibitem{gwtc1}
{\scshape LIGO Scientific, VIRGO} collaboration, \emph{{GWTC-2.1: Deep extended catalog of compact binary coalescences observed by LIGO and Virgo during the first half of the third observing run}}, \href{https://doi.org/10.1103/PhysRevD.109.022001}{\emph{Phys. Rev. D} {\bfseries 109} (2024) 022001} [\href{https://arxiv.org/abs/2108.01045}{{\ttfamily 2108.01045}}].

\bibitem{gwtc3-1}
{\scshape LIGO Scientific, Virgo} collaboration, \emph{{GWTC-2: Compact Binary Coalescences Observed by LIGO and Virgo During the First Half of the Third Observing Run}}, \href{https://doi.org/10.1103/PhysRevX.11.021053}{\emph{Phys. Rev. X} {\bfseries 11} (2021) 021053} [\href{https://arxiv.org/abs/2010.14527}{{\ttfamily 2010.14527}}].

\bibitem{gwtc3}
{\scshape KAGRA, VIRGO, LIGO Scientific} collaboration, \emph{{GWTC-3: Compact Binary Coalescences Observed by LIGO and Virgo during the Second Part of the Third Observing Run}}, \href{https://doi.org/10.1103/PhysRevX.13.041039}{\emph{Phys. Rev. X} {\bfseries 13} (2023) 041039} [\href{https://arxiv.org/abs/2111.03606}{{\ttfamily 2111.03606}}].

\bibitem{BBH_fc}
M.~{Mapelli}, \emph{{Astrophysics of stellar black holes}}, \href{https://doi.org/10.48550/arXiv.1809.09130}{\emph{arXiv e-prints} (2018) arXiv:1809.09130} [\href{https://arxiv.org/abs/1809.09130}{{\ttfamily 1809.09130}}].

\bibitem{Mandel}
I.~Mandel and A.~Farmer, \emph{{Merging stellar-mass binary black holes}}, \href{https://doi.org/10.1016/j.physrep.2022.01.003}{\emph{Phys. Rept.} {\bfseries 955} (2022) 1} [\href{https://arxiv.org/abs/1806.05820}{{\ttfamily 1806.05820}}].

\bibitem{Romero-Shaw1}
I.M.~Romero-Shaw, P.D.~Lasky and E.~Thrane, \emph{{Searching for Eccentricity: Signatures of Dynamical Formation in the First Gravitational-Wave Transient Catalogue of LIGO and Virgo}}, \href{https://doi.org/10.1093/mnras/stz2996}{\emph{Mon. Not. Roy. Astron. Soc.} {\bfseries 490} (2019) 5210} [\href{https://arxiv.org/abs/1909.05466}{{\ttfamily 1909.05466}}].

\bibitem{peter}
P.C.~Peters, \emph{Gravitational radiation and the motion of two point masses}, \href{https://doi.org/10.1103/PhysRev.136.B1224}{\emph{Phys. Rev.} {\bfseries 136} (1964) B1224}.

\bibitem{Celoria}
M.~Celoria, R.~Oliveri, A.~Sesana and M.~Mapelli, \emph{{Lecture notes on black hole binary astrophysics}},  7, 2018 [\href{https://arxiv.org/abs/1807.11489}{{\ttfamily 1807.11489}}].

\bibitem{DiCarlo}
U.N.~Di~Carlo, N.~Giacobbo, M.~Mapelli, M.~Pasquato, M.~Spera, L.~Wang et~al., \emph{{Merging black holes in young star clusters}}, \href{https://doi.org/10.1093/mnras/stz1453}{\emph{Mon. Not. Roy. Astron. Soc.} {\bfseries 487} (2019) 2947} [\href{https://arxiv.org/abs/1901.00863}{{\ttfamily 1901.00863}}].

\bibitem{Banerjee}
S.~Banerjee, \emph{{Stellar-mass black holes in young massive and open stellar clusters and their role in gravitational-wave generation \textendash{} II}}, \href{https://doi.org/10.1093/mnras/stx2347}{\emph{Mon. Not. Roy. Astron. Soc.} {\bfseries 473} (2018) 909} [\href{https://arxiv.org/abs/1707.00922}{{\ttfamily 1707.00922}}].

\bibitem{Rodriguez2}
C.L.~Rodriguez, S.~Chatterjee and F.A.~Rasio, \emph{{Binary Black Hole Mergers from Globular Clusters: Masses, Merger Rates, and the Impact of Stellar Evolution}}, \href{https://doi.org/10.1103/PhysRevD.93.084029}{\emph{Phys. Rev. D} {\bfseries 93} (2016) 084029} [\href{https://arxiv.org/abs/1602.02444}{{\ttfamily 1602.02444}}].

\bibitem{Rodriguez}
C.L.~Rodriguez, P.~Amaro-Seoane, S.~Chatterjee and F.A.~Rasio, \emph{{Post-Newtonian Dynamics in Dense Star Clusters: Highly-Eccentric, Highly-Spinning, and Repeated Binary Black Hole Mergers}}, \href{https://doi.org/10.1103/PhysRevLett.120.151101}{\emph{Phys. Rev. Lett.} {\bfseries 120} (2018) 151101} [\href{https://arxiv.org/abs/1712.04937}{{\ttfamily 1712.04937}}].

\bibitem{Samsing2}
J.~Samsing, \emph{{Eccentric Black Hole Mergers Forming in Globular Clusters}}, \href{https://doi.org/10.1103/PhysRevD.97.103014}{\emph{Phys. Rev. D} {\bfseries 97} (2018) 103014} [\href{https://arxiv.org/abs/1711.07452}{{\ttfamily 1711.07452}}].

\bibitem{Zevin}
M.~Zevin, I.M.~Romero-Shaw, K.~Kremer, E.~Thrane and P.D.~Lasky, \emph{{Implications of Eccentric Observations on Binary Black Hole Formation Channels}}, \href{https://doi.org/10.3847/2041-8213/ac32dc}{\emph{Astrophys. J. Lett.} {\bfseries 921} (2021) L43} [\href{https://arxiv.org/abs/2106.09042}{{\ttfamily 2106.09042}}].

\bibitem{Samsing}
J.~Samsing, I.~Bartos, D.J.~D'Orazio, Z.~Haiman, B.~Kocsis, N.W.C.~Leigh et~al., \emph{{AGN as potential factories for eccentric black hole mergers}}, \href{https://doi.org/10.1038/s41586-021-04333-1}{\emph{Nature} {\bfseries 603} (2022) 237} [\href{https://arxiv.org/abs/2010.09765}{{\ttfamily 2010.09765}}].

\bibitem{Tagawa}
H.~Tagawa, B.~Kocsis, Z.~Haiman, I.~Bartos, K.~Omukai and J.~Samsing, \emph{{Eccentric Black Hole Mergers in Active Galactic Nuclei}}, \href{https://doi.org/10.3847/2041-8213/abd4d3}{\emph{Astrophys. J. Lett.} {\bfseries 907} (2021) L20} [\href{https://arxiv.org/abs/2010.10526}{{\ttfamily 2010.10526}}].

\bibitem{gw19}
{\scshape LIGO Scientific, Virgo} collaboration, \emph{{GW190521: A Binary Black Hole Merger with a Total Mass of $150 M_{\odot}$}}, \href{https://doi.org/10.1103/PhysRevLett.125.101102}{\emph{Phys. Rev. Lett.} {\bfseries 125} (2020) 101102} [\href{https://arxiv.org/abs/2009.01075}{{\ttfamily 2009.01075}}].

\bibitem{gw19-2}
{\scshape LIGO Scientific, Virgo} collaboration, \emph{{Properties and Astrophysical Implications of the 150 M$_\odot$ Binary Black Hole Merger GW190521}}, \href{https://doi.org/10.3847/2041-8213/aba493}{\emph{Astrophys. J. Lett.} {\bfseries 900} (2020) L13} [\href{https://arxiv.org/abs/2009.01190}{{\ttfamily 2009.01190}}].

\bibitem{gw19-3}
I.M.~Romero-Shaw, P.D.~Lasky, E.~Thrane and J.C.~Bustillo, \emph{{GW190521: orbital eccentricity and signatures of dynamical formation in a binary black hole merger signal}}, \href{https://doi.org/10.3847/2041-8213/abbe26}{\emph{Astrophys. J. Lett.} {\bfseries 903} (2020) L5} [\href{https://arxiv.org/abs/2009.04771}{{\ttfamily 2009.04771}}].

\bibitem{gw19-4}
V.~Gayathri, J.~Healy, J.~Lange, B.~O'Brien, M.~Szczepanczyk, I.~Bartos et~al., \emph{{Eccentricity estimate for black hole mergers with numerical relativity simulations}}, \href{https://doi.org/10.1038/s41550-021-01568-w}{\emph{Nature Astron.} {\bfseries 6} (2022) 344} [\href{https://arxiv.org/abs/2009.05461}{{\ttfamily 2009.05461}}].

\bibitem{Romeroecc2}
I.M.~Romero-Shaw, P.D.~Lasky and E.~Thrane, \emph{{Signs of Eccentricity in Two Gravitational-wave Signals May Indicate a Subpopulation of Dynamically Assembled Binary Black Holes}}, \href{https://doi.org/10.3847/2041-8213/ac3138}{\emph{Astrophys. J. Lett.} {\bfseries 921} (2021) L31} [\href{https://arxiv.org/abs/2108.01284}{{\ttfamily 2108.01284}}].

\bibitem{Romeroecc}
I.M.~Romero-Shaw, P.D.~Lasky and E.~Thrane, \emph{{Four Eccentric Mergers Increase the Evidence that LIGO\textendash{}Virgo\textendash{}KAGRA\textquoteright{}s Binary Black Holes Form Dynamically}}, \href{https://doi.org/10.3847/1538-4357/ac9798}{\emph{Astrophys. J.} {\bfseries 940} (2022) 171} [\href{https://arxiv.org/abs/2206.14695}{{\ttfamily 2206.14695}}].

\bibitem{Iglesias}
H.L.~Iglesias et~al., \emph{{Eccentricity Estimation for Five Binary Black Hole Mergers with Higher-order Gravitational-wave Modes}}, \href{https://doi.org/10.3847/1538-4357/ad5ff6}{\emph{Astrophys. J.} {\bfseries 972} (2024) 65} [\href{https://arxiv.org/abs/2208.01766}{{\ttfamily 2208.01766}}].

\bibitem{Gupte}
N.~Gupte et~al., \emph{{Evidence for eccentricity in the population of binary black holes observed by LIGO-Virgo-KAGRA}},  \href{https://arxiv.org/abs/2404.14286}{{\ttfamily 2404.14286}}.

\bibitem{Morras}
G.~{Morras}, G.~{Pratten} and P.~{Schmidt}, \emph{{Orbital eccentricity in a neutron star - black hole binary}}, \href{https://doi.org/10.48550/arXiv.2503.15393}{\emph{arXiv e-prints} (2025) arXiv:2503.15393} [\href{https://arxiv.org/abs/2503.15393}{{\ttfamily 2503.15393}}].

\bibitem{pycbcPE}
C.M.~{Biwer}, C.D.~{Capano}, S.~{De}, M.~{Cabero}, D.A.~{Brown}, A.H.~{Nitz} et~al., \emph{{PyCBC Inference: A Python-based Parameter Estimation Toolkit for Compact Binary Coalescence Signal}}, \href{https://doi.org/10.1088/1538-3873/aaef0b}{\emph{\pasp} {\bfseries 131} (2019) 024503} [\href{https://arxiv.org/abs/1807.10312}{{\ttfamily 1807.10312}}].

\bibitem{bilby}
G.~{Ashton}, M.~{H{\"u}bner}, P.D.~{Lasky}, C.~{Talbot}, K.~{Ackley}, S.~{Biscoveanu} et~al., \emph{{BILBY: A User-friendly Bayesian Inference Library for Gravitational-wave Astronomy}}, \href{https://doi.org/10.3847/1538-4365/ab06fc}{\emph{\apjs} {\bfseries 241} (2019) 27} [\href{https://arxiv.org/abs/1811.02042}{{\ttfamily 1811.02042}}].

\bibitem{Divyajyoti}
Divyajyoti, S.~Kumar, S.~Tibrewal, I.M.~Romero-Shaw and C.K.~Mishra, \emph{{Blind spots and biases: The dangers of ignoring eccentricity in gravitational-wave signals from binary black holes}}, \href{https://doi.org/10.1103/PhysRevD.109.043037}{\emph{Phys. Rev. D} {\bfseries 109} (2024) 043037} [\href{https://arxiv.org/abs/2309.16638}{{\ttfamily 2309.16638}}].

\bibitem{Favata}
M.~Favata, C.~Kim, K.G.~Arun, J.~Kim and H.W.~Lee, \emph{{Constraining the orbital eccentricity of inspiralling compact binary systems with Advanced LIGO}}, \href{https://doi.org/10.1103/PhysRevD.105.023003}{\emph{Phys. Rev. D} {\bfseries 105} (2022) 023003} [\href{https://arxiv.org/abs/2108.05861}{{\ttfamily 2108.05861}}].

\bibitem{Brown}
D.A.~Brown and P.J.~Zimmerman, \emph{{The Effect of Eccentricity on Searches for Gravitational-Waves from Coalescing Compact Binaries in Ground-based Detectors}}, \href{https://doi.org/10.1103/PhysRevD.81.024007}{\emph{Phys. Rev. D} {\bfseries 81} (2010) 024007} [\href{https://arxiv.org/abs/0909.0066}{{\ttfamily 0909.0066}}].

\bibitem{Huerta}
E.A.~Huerta and D.A.~Brown, \emph{{Effect of eccentricity on binary neutron star searches in Advanced LIGO}}, \href{https://doi.org/10.1103/PhysRevD.87.127501}{\emph{Phys. Rev. D} {\bfseries 87} (2013) 127501} [\href{https://arxiv.org/abs/1301.1895}{{\ttfamily 1301.1895}}].

\bibitem{Huerta2}
E.A.~Huerta et~al., \emph{{Complete waveform model for compact binaries on eccentric orbits}}, \href{https://doi.org/10.1103/PhysRevD.95.024038}{\emph{Phys. Rev. D} {\bfseries 95} (2017) 024038} [\href{https://arxiv.org/abs/1609.05933}{{\ttfamily 1609.05933}}].

\bibitem{detection1}
D.~{George} and E.A.~{Huerta}, \emph{{Deep Learning for real-time gravitational wave detection and parameter estimation: Results with Advanced LIGO data}}, \href{https://doi.org/10.1016/j.physletb.2017.12.053}{\emph{Physics Letters B} {\bfseries 778} (2018) 64} [\href{https://arxiv.org/abs/1711.03121}{{\ttfamily 1711.03121}}].

\bibitem{detection2}
H.~{Gabbard}, M.~{Williams}, F.~{Hayes} and C.~{Messenger}, \emph{{Matching Matched Filtering with Deep Networks for Gravitational-Wave Astronomy}}, \href{https://doi.org/10.1103/PhysRevLett.120.141103}{\emph{\prl} {\bfseries 120} (2018) 141103} [\href{https://arxiv.org/abs/1712.06041}{{\ttfamily 1712.06041}}].

\bibitem{detection3}
P.~{Nousi}, A.E.~{Koloniari}, N.~{Passalis}, P.~{Iosif}, N.~{Stergioulas} and A.~{Tefas}, \emph{{Deep residual networks for gravitational wave detection}}, \href{https://doi.org/10.1103/PhysRevD.108.024022}{\emph{\prd} {\bfseries 108} (2023) 024022} [\href{https://arxiv.org/abs/2211.01520}{{\ttfamily 2211.01520}}].

\bibitem{glitch}
C.~{Chatterjee}, A.~{Petulante}, K.~{Jani}, J.~{Spencer-Smith}, Y.~{Hu}, R.~{Lau} et~al., \emph{{Pre-trained Audio Transformer as a Foundational AI Tool for Gravitational Waves}}, \href{https://doi.org/10.48550/arXiv.2412.20789}{\emph{arXiv e-prints} (2024) arXiv:2412.20789} [\href{https://arxiv.org/abs/2412.20789}{{\ttfamily 2412.20789}}].

\bibitem{PE2}
M.~{Dax}, S.R.~{Green}, J.~{Gair}, J.H.~{Macke}, A.~{Buonanno} and B.~{Sch{\"o}lkopf}, \emph{{Real-Time Gravitational Wave Science with Neural Posterior Estimation}}, \href{https://doi.org/10.1103/PhysRevLett.127.241103}{\emph{\prl} {\bfseries 127} (2021) 241103} [\href{https://arxiv.org/abs/2106.12594}{{\ttfamily 2106.12594}}].

\bibitem{PE1}
H.~{Gabbard}, C.~{Messenger}, I.S.~{Heng}, F.~{Tonolini} and R.~{Murray-Smith}, \emph{{Bayesian parameter estimation using conditional variational autoencoders for gravitational-wave astronomy}}, \href{https://doi.org/10.1038/s41567-021-01425-7}{\emph{Nature Physics} {\bfseries 18} (2022) 112} [\href{https://arxiv.org/abs/1909.06296}{{\ttfamily 1909.06296}}].

\bibitem{PE3}
M.~{Dax}, S.R.~{Green}, J.~{Gair}, N.~{Gupte}, M.~{P{\"u}rrer}, V.~{Raymond} et~al., \emph{{Real-time inference for binary neutron star mergers using machine learning}}, \href{https://doi.org/10.1038/s41586-025-08593-z}{\emph{\nat} {\bfseries 639} (2025) 49} [\href{https://arxiv.org/abs/2407.09602}{{\ttfamily 2407.09602}}].

\bibitem{other}
E.~{Cuoco}, M.~{Cavagli{\`a}}, I.S.~{Heng}, D.~{Keitel} and C.~{Messenger}, \emph{{Applications of machine learning in gravitational-wave research with current interferometric detectors: Applications of machine learning in gravitational-wave...}}, \href{https://doi.org/10.1007/s41114-024-00055-8}{\emph{Living Reviews in Relativity} {\bfseries 28} (2025) 2} [\href{https://arxiv.org/abs/2412.15046}{{\ttfamily 2412.15046}}].

\bibitem{transformer1}
A.~{Vaswani}, N.~{Shazeer}, N.~{Parmar}, J.~{Uszkoreit}, L.~{Jones}, A.N.~{Gomez} et~al., \emph{{Attention Is All You Need}}, \href{https://doi.org/10.48550/arXiv.1706.03762}{\emph{arXiv e-prints} (2017) arXiv:1706.03762} [\href{https://arxiv.org/abs/1706.03762}{{\ttfamily 1706.03762}}].

\bibitem{GPT4}
{OpenAI}, J.~{Achiam}, S.~{Adler}, S.~{Agarwal}, L.~{Ahmad}, I.~{Akkaya} et~al., \emph{{GPT-4 Technical Report}}, \href{https://doi.org/10.48550/arXiv.2303.08774}{\emph{arXiv e-prints} (2023) arXiv:2303.08774} [\href{https://arxiv.org/abs/2303.08774}{{\ttfamily 2303.08774}}].

\bibitem{transformer2}
A.~{Dosovitskiy}, L.~{Beyer}, A.~{Kolesnikov}, D.~{Weissenborn}, X.~{Zhai}, T.~{Unterthiner} et~al., \emph{{An Image is Worth 16x16 Words: Transformers for Image Recognition at Scale}}, \href{https://doi.org/10.48550/arXiv.2010.11929}{\emph{arXiv e-prints} (2020) arXiv:2010.11929} [\href{https://arxiv.org/abs/2010.11929}{{\ttfamily 2010.11929}}].

\bibitem{transformer3}
Q.~{Wen}, T.~{Zhou}, C.~{Zhang}, W.~{Chen}, Z.~{Ma}, J.~{Yan} et~al., \emph{{Transformers in Time Series: A Survey}}, \href{https://doi.org/10.48550/arXiv.2202.07125}{\emph{arXiv e-prints} (2022) arXiv:2202.07125} [\href{https://arxiv.org/abs/2202.07125}{{\ttfamily 2202.07125}}].

\bibitem{transformer5}
S.Y.~{Hwang}, C.G.~{Sabiu}, I.~{Park} and S.E.~{Hong}, \emph{{The universe is worth {}64$^{3}$ pixels: convolution neural network and vision transformers for cosmology}}, \href{https://doi.org/10.1088/1475-7516/2023/11/075}{\emph{\jcap} {\bfseries 2023} (2023) 075} [\href{https://arxiv.org/abs/2304.08192}{{\ttfamily 2304.08192}}].

\bibitem{transformer4}
T.~{Allam} and J.D.~{McEwen}, \emph{{Paying attention to astronomical transients: introducing the time-series transformer for photometric classification}}, \href{https://doi.org/10.1093/rasti/rzad046}{\emph{RAS Techniques and Instruments} {\bfseries 3} (2024) 209} [\href{https://arxiv.org/abs/2105.06178}{{\ttfamily 2105.06178}}].

\bibitem{transformer6}
O.M.~{Boersma}, E.H.~{Ayache} and J.~{van Leeuwen}, \emph{{Transformer models for astrophysical time series and the GRB prompt-afterglow relation}}, \href{https://doi.org/10.1093/rasti/rzae026}{\emph{RAS Techniques and Instruments} {\bfseries 3} (2024) 472} [\href{https://arxiv.org/abs/2406.12515}{{\ttfamily 2406.12515}}].

\bibitem{LSTM}
S.~Hochreiter and J.~Schmidhuber, \emph{Long short-term memory}, \href{https://doi.org/10.1162/neco.1997.9.8.1735}{\emph{Neural Computation} {\bfseries 9} (1997) 1735}.

\bibitem{external}
M.-H.~{Guo}, Z.-N.~{Liu}, T.-J.~{Mu} and S.-M.~{Hu}, \emph{{Beyond Self-attention: External Attention using Two Linear Layers for Visual Tasks}}, \href{https://doi.org/10.48550/arXiv.2105.02358}{\emph{arXiv e-prints} (2021) arXiv:2105.02358} [\href{https://arxiv.org/abs/2105.02358}{{\ttfamily 2105.02358}}].

\bibitem{formula}
G.~{Wang}, J.~{Wang}, C.~{Zhou}, W.~{Ding}, H.~{Zeng}, T.~{Li} et~al., \emph{{Decoding Emotions: Unveiling Facial Expressions through Acoustic Sensing with Contrastive Attention}}, \href{https://doi.org/10.48550/arXiv.2410.12811}{\emph{arXiv e-prints} (2024) arXiv:2410.12811} [\href{https://arxiv.org/abs/2410.12811}{{\ttfamily 2410.12811}}].

\bibitem{DGI}
T.~{Damour}, A.~{Gopakumar} and B.R.~{Iyer}, \emph{{Phasing of gravitational waves from inspiralling eccentric binaries}}, \href{https://doi.org/10.1103/PhysRevD.70.064028}{\emph{\prd} {\bfseries 70} (2004) 064028} [\href{https://arxiv.org/abs/gr-qc/0404128}{{\ttfamily gr-qc/0404128}}].

\bibitem{Gamboa}
A.~{Gamboa}, A.~{Buonanno}, R.~{Enficiaud}, M.~{Khalil}, A.~{Ramos-Buades}, L.~{Pompili} et~al., \emph{{Accurate waveforms for eccentric, aligned-spin binary black holes: The multipolar effective-one-body model SEOBNRv5EHM}}, \href{https://doi.org/10.48550/arXiv.2412.12823}{\emph{arXiv e-prints} (2024) arXiv:2412.12823} [\href{https://arxiv.org/abs/2412.12823}{{\ttfamily 2412.12823}}].

\bibitem{EOB5}
A.~{Gamboa}, M.~{Khalil} and A.~{Buonanno}, \emph{{Third post-Newtonian dynamics for eccentric orbits and aligned spins in the effective-one-body waveform model SEOBNRv5EHM}}, \href{https://doi.org/10.48550/arXiv.2412.12831}{\emph{arXiv e-prints} (2024) arXiv:2412.12831} [\href{https://arxiv.org/abs/2412.12831}{{\ttfamily 2412.12831}}].

\bibitem{EOB1}
Z.~{Cao} and W.-B.~{Han}, \emph{{Waveform model for an eccentric binary black hole based on the effective-one-body-numerical-relativity formalism}}, \href{https://doi.org/10.1103/PhysRevD.96.044028}{\emph{\prd} {\bfseries 96} (2017) 044028} [\href{https://arxiv.org/abs/1708.00166}{{\ttfamily 1708.00166}}].

\bibitem{pys}
D.P.~Mihaylov, S.~Ossokine, A.~Buonanno, H.~Estelles, L.~Pompili, M.~P\"urrer et~al., \emph{{pySEOBNR: a software package for the next generation of effective-one-body multipolar waveform models}},  \href{https://arxiv.org/abs/2303.18203}{{\ttfamily 2303.18203}}.

\bibitem{lal1}
K.~{Wette}, \emph{{SWIGLAL: Python and Octave interfaces to the LALSuite gravitational-wave data analysis libraries}}, \href{https://doi.org/10.1016/j.softx.2020.100634}{\emph{SoftwareX} {\bfseries 12} (2020) 100634} [\href{https://arxiv.org/abs/2012.09552}{{\ttfamily 2012.09552}}].

\bibitem{lal2}
{LIGO Scientific Collaboration}, {Virgo Collaboration} and {KAGRA Collaboration}, ``{LVK} {A}lgorithm {L}ibrary - {LALS}uite.'' Free software (GPL), 2018.
\newblock 10.7935/GT1W-FZ16.

\bibitem{EOB4}
M.~{Khalil}, A.~{Buonanno}, J.~{Steinhoff} and J.~{Vines}, \emph{{Radiation-reaction force and multipolar waveforms for eccentric, spin-aligned binaries in the effective-one-body formalism}}, \href{https://doi.org/10.1103/PhysRevD.104.024046}{\emph{\prd} {\bfseries 104} (2021) 024046} [\href{https://arxiv.org/abs/2104.11705}{{\ttfamily 2104.11705}}].

\bibitem{EOB4-2}
A.~{Ramos-Buades}, A.~{Buonanno}, M.~{Khalil} and S.~{Ossokine}, \emph{{Effective-one-body multipolar waveforms for eccentric binary black holes with nonprecessing spins}}, \href{https://doi.org/10.1103/PhysRevD.105.044035}{\emph{\prd} {\bfseries 105} (2022) 044035} [\href{https://arxiv.org/abs/2112.06952}{{\ttfamily 2112.06952}}].

\bibitem{adh}
A.~{Ravichandran}, A.~{Vijaykumar}, S.J.~{Kapadia} and P.~{Kumar}, \emph{{Rapid Identification and Classification of Eccentric Gravitational Wave Inspirals with Machine Learning}}, \href{https://doi.org/10.48550/arXiv.2302.00666}{\emph{arXiv e-prints} (2023) arXiv:2302.00666} [\href{https://arxiv.org/abs/2302.00666}{{\ttfamily 2302.00666}}].

\bibitem{pycbc}
A.~Nitz, I.~Harry, D.~Brown, C.M.~Biwer, J.~Willis, T.D.~Canton et~al., \emph{gwastro/pycbc: v2.3.3 release of pycbc},  Jan., 2024.
\newblock 10.5281/zenodo.10473621.

\bibitem{em}
B.G.~{Patterson}, S.M.~{Tomson} and S.~{Fairhurst}, \emph{{Identifying eccentricity in binary black hole mergers using a harmonic decomposition of the gravitational waveform}}, \href{https://doi.org/10.1103/PhysRevD.111.044073}{\emph{\prd} {\bfseries 111} (2025) 044073} [\href{https://arxiv.org/abs/2411.04187}{{\ttfamily 2411.04187}}].

\bibitem{em1}
M.~{Favata}, C.~{Kim}, K.G.~{Arun}, J.~{Kim} and H.W.~{Lee}, \emph{{Constraining the orbital eccentricity of inspiralling compact binary systems with Advanced LIGO}}, \href{https://doi.org/10.1103/PhysRevD.105.023003}{\emph{\prd} {\bfseries 105} (2022) 023003} [\href{https://arxiv.org/abs/2108.05861}{{\ttfamily 2108.05861}}].

\bibitem{Lower}
M.E.~Lower, E.~Thrane, P.D.~Lasky and R.~Smith, \emph{{Measuring eccentricity in binary black hole inspirals with gravitational waves}}, \href{https://doi.org/10.1103/PhysRevD.98.083028}{\emph{Phys. Rev. D} {\bfseries 98} (2018) 083028} [\href{https://arxiv.org/abs/1806.05350}{{\ttfamily 1806.05350}}].

\bibitem{mbias}
Y.-S.~{Ting}, \emph{{Why Machine Learning Models Systematically Underestimate Extreme Values}}, \href{https://doi.org/10.48550/arXiv.2412.05806}{\emph{arXiv e-prints} (2024) arXiv:2412.05806} [\href{https://arxiv.org/abs/2412.05806}{{\ttfamily 2412.05806}}].

\end{thebibliography}\endgroup

\appendix
 
\end{document}